\documentclass[aps,pra,superscriptaddress,twocolumn,amsmath,amsfonts,amssymb,floatfix,nofootinbib]{revtex4}
\usepackage{enumerate}
\usepackage{mathtools}
\usepackage{amsmath}
\usepackage{graphicx}
\usepackage{sidecap}
\usepackage[normalem]{ulem}
\usepackage{color}
\usepackage{enumerate}
\usepackage{bm}

\usepackage[utf8]{inputenc}
\usepackage{amsfonts}
\usepackage{amssymb}
\usepackage{enumitem}

\makeatletter

\date{}

\providecommand{\tabularnewline}{\\}
\makeatother

\begin{document}

\title{Nonclassicality in off-resonant Raman process}
%

\author{Kishore Thapliyal} \thanks{Email:tkishore36@yahoo.com}
\affiliation{RCPTM, Joint Laboratory of Optics of Palacky University
and Institute of Physics of Academy of Science of the Czech Republic,
Faculty of Science, Palacky University, 17. listopadu 12, 771 46 Olomouc,
Czech Republic}

\author{Jan Pe{\v r}ina} 
\affiliation{RCPTM, Joint Laboratory of Optics of Palacky University
and Institute of Physics of Academy of Science of the Czech Republic,
Faculty of Science, Palacky University, 17. listopadu 12, 771 46 Olomouc,
Czech Republic}
\affiliation{Department of Optics, Faculty of Science, Palacky University, 17. listopadu 12, 771 46 Olomouc, Czech Republic}

\begin{abstract}
Raman process is used for generation of nonclassical states and subsequent applications due to its simplicity. In view of the experimental feasibility of the off-resonant Raman process at single photon level useful in long distance quantum communication and quantum memory, we perform here a detailed study of possibility of generation of nonclassical states under off-resonance conditions. Specifically, we have studied
nonclassicality in the off-resonant Raman process with the help of a characteristic
function written using more general solution than the conventional short-time solution under complete quantum treatment and established the significance of frequency detuning parameter in generation of nonclassical states. The 
obtained characteristic function remains
Gaussian and reveals that larger values of frequency detuning parameters are favorable to induce nonclassicality in some cases. The single- and two-mode squeezing and antibunching as well as intermodal entanglement are reported in terms of experimentally accessible quantum noise fluctuations considering either phonon mode coherent or chaotic. Subsequently, the joint photon number and integrated intensity distributions are studied to analyze the performance of pair generation in Stokes-phonon and pump-phonon modes. The present study discusses the role of resonance conditions in generation of nonclassical states and establishes that experimentally controllable detuning parameter can be used to probe and enhance the generated nonclassicality.
\end{abstract}

\maketitle

\section{Introduction}

The role of nonclassical states is inevitably at the core of the recent growth in quantum information processing and technology. Moreover, the future development of the field depends upon our theoretical understanding of nonclassical phenomena and experimental availability of the physical systems and processes able to generate such states. Specifically, the quantum states with negative Glauber-Sudarshan $P$ function \cite{glauber1963coherent,sudarshan1963equivalence} are known as nonclassical states due to unavailability of a classical counterpart for such states. These nonclassical states can be generated from nonlinear optical couplers \cite{thapliyal2014higher,thapliyal2014nonclassical}, Bose-Einstein condensates \cite{zeng1995nonclassical,giri2017nonclassicality}, optical \cite{naikoo2018probing,naikoo2018quantum} and  optomechanical \cite{brooks2012non,alam2017lower} cavity systems as well as nonlinear optical processes, such as parametric down-conversion \cite{wu1986generation,perina2011joint}, four-wave mixing \cite{slusher1985observation}, Raman and hyper-Raman scattering \cite{perina1992quantum,thapliyal2017nonclassicality} (see \cite{perina1991quantum} for review). 
A few examples of nonclassical states useful in quantum information processing are single photon source (which is an antibunched state) \cite{bennett1984quantum}, squeezed \cite{hillery2000quantum}, entangled \cite{ekert1991quantum}, steerable \cite{branciard2012one}, and Bell nonlocal \cite{acin2006bell} states particularly used for quantum cryptography and quantum random number generation \cite{herrero2017quantum}.  Entangled states have applications in quantum metrology \cite{giovannetti2011advances}, teleportation \cite{bennett1993teleporting}, and densecoding \cite{bennett1992communication}, too.  Squeezed states are also used in quantum metrology \cite{giovannetti2011advances}, teleportation \cite{braunstein1998teleportation}, and detection of gravitational wave \cite{abbott2016observation,abbott2016gw151226}. 

These applications of different nonclassical features have motivated in a set of theoretical (\cite{miranowicz1994quantum,sen2005squeezed,sen2007quantum,sen2008amplitude,sen2011sub,pathak2013nonclassicality,sen2013intermodal,giri2016higher}
and references therein, Section 10.4 of \cite{perina1991quantum}) and experimental \cite{meekhof1996generation,chen2006deterministic,matsukevich2004quantum,lee2012macroscopic,kasperczyk2016temporal} studies of nonclassicality in Raman scattering. Specifically, quantum theory of Raman effects was developed  with frequency detuning \cite{walls1970quantum}; squeezing
\cite{sen2005squeezed}, antibunching \cite{sen2007quantum}, amplitude powered squeezing \cite{sen2008amplitude}, sub-shot noise \cite{sen2011sub}, lower-order entanglement \cite{sen2013intermodal}, and higher-order entanglement \cite{giri2016higher} in spontaneous and stimulated Raman processes are reported; study on quantum statistics with squeezed light and pump depletion has also been performed \cite{perina1992quantum}. Experimental generation of Fock and squeezed states \cite{meekhof1996generation}, a single photon source \cite{chen2006deterministic}, quantum state transfer between matter and light \cite{matsukevich2004quantum}, nonclassical phonon states in diamond \cite{lee2012macroscopic}, and Stokes-anti-Stokes correlation in scattering from water \cite{kasperczyk2016temporal}  are reported using Raman processes. Further, long distance quantum communication using off-resonant Raman process is proposed \cite{duan2001long}, which motivated a set of experiments for the generation of photon pairs \cite{kuzmich2003generation}, photon-phonon correlated pairs \cite{riedinger2016non}, and quantum memory \cite{dou2018broadband,ding2018raman}. More recently, Raman processes are also studied using the theoretical models of Cooper pair \cite{saraiva2017photonic} and optomechanical \cite{roelli2016molecular} systems.

A complete quantum mechanical treatment of the Raman process is complicated and is thus often simplified by assuming strong classical pump leading to a set of linear coupled Heisenberg's equation of motion \cite{pieczonkova1981statistical,perina1991quantum}. Additionally, time evolution of each mode valid only in the short-time domain is reported under fully quantum treatment \cite{perina1991quantum}. In both these cases, frequency matching conditions are assumed. In contrast, energy (frequency) mismatch is also introduced in \cite{walls1970quantum}. A perturbative technique to obtain time evolution of operators in a more general form than that of short-time solution is also used recently for Raman process (\cite{sen2013intermodal,giri2016higher} and references therein). The solution obtained using this technique is more general, as the short-time solution reported in \cite{perina1991quantum} can be obtained by neglecting terms beyond quadratic, and intrinsically comes in the form of frequency detunings in Stokes and anti-Stokes generation. The frequency detuning is experimentally a relevant condition as the
Stokes and anti-Stokes generations are expected to be high for the resonant Raman conditions, but which are often hard to achieve in an experiment. In some experiments
\cite{duan2001long,kuzmich2003generation,riedinger2016non,dou2018broadband,ding2018raman},
large detuning has its own advantages. The present solution may be useful in highlighting such benefits in the generation of nonclassicality in the off-resonant Raman processes. However, in the past, no such attempt has been made. Therefore, here we set a twofold motivation. Firstly, whether the characteristic
function obtained using a more general solution than short-time solution \cite{sen2013intermodal,thapliyal2017nonclassicality}
remains Gaussian or not. In the past, it has been observed that a
more general Raman process--hyper-Raman process (\cite{thapliyal2017nonclassicality} and references therein)--has non-Gaussian
characteristic function also for the short-time solution \cite{perinova1979quantum}. Secondly, to probe nonclassicality in the cases when short-time solution failed to detect it using 
a more general solution \cite{thapliyal2017nonclassicality}.

The rest of the paper is organized as follows. Section \ref{sec:sys} is devoted to a brief discussion of the Hamiltonian describing the Raman process and corresponding characteristic function. Subsequently the obtained characteristic function assuming all modes initially coherent is used to study a set of nonclassical features in Section \ref{sec:coh}. In the next section, characteristic function is obtained considering phonon mode chaotic, and a brief discussion of nonclassical features follows. Thereafter, joint photon-phonon number and integrated intensity distributions are obtained in Section \ref{sec:joint}, which are further used to obtain difference and conditional number distributions in Section \ref{sec:cond} before concluding in Section \ref{sec:Conclusion}. 

\section{Raman process and the characteristic function \label{sec:sys}}

The Hamiltonian for the stimulated and spontaneous Raman processes
in a complete quantum mechanical description is \cite{perina1991quantum}

\begin{equation}
\begin{array}{l}
H  =  \underset{j=L,S,A,V}{\sum}\hbar\omega_{j}a_{j}^{\dagger}a_{j}-\left(ga_{L}a_{S}^{\dagger}a_{V}^{\dagger}+\chi^{*}a_{L}a_{V}a_{A}^{\dagger}+\textrm{H.c.}\right),\end{array}\label{eq:Ham}
\end{equation}
where H.c. stands for the Hermitian conjugate. The subscripts $L,\,S,\,V,$
and $A$ correspond to the laser (pump), Stokes, vibration (phonon),
and anti-Stokes modes, respectively. The annihilation (creation) operator
$a_{j}\,(a_{j}^{\dagger})$ corresponds to the $j$th mode with frequency
$\omega_{j}$. Also, $g$ and $\chi$ are Stokes and anti-Stokes coupling
constants, respectively. The Sen-Mandal perturbative solution of Hamiltonian
(\ref{eq:Ham}) is reported in (\cite{thapliyal2017nonclassicality} and references therein),
which is 
\begin{equation}
\begin{array}{lcl}
a_{L}(t) & = & f_{1}a_{L}(0)+f_{2}a_{S}(0)a_{V}(0)+f_{3}a_{V}^{\dagger}(0)a_{A}(0)\\
 & + & f_{4}a_{L}^{\dagger}(0)a_{S}(0)a_{A}(0)+f_{5}a_{L}(0)a_{S}(0)a_{S}^{\dagger}(0)\\
 & + & f_{6}a_{L}(0)a_{V}^{\dagger}(0)a_{V}(0)+f_{7}a_{L}(0)a_{V}^{\dagger}(0)a_{V}(0)\\
 & + & f_{8}a_{L}(0)a_{A}^{\dagger}(0)a_{A}(0),\\
a_{S}(t) & = & g_{1}a_{S}(0)+g_{2}a_{L}(0)a_{V}^{\dagger}(0)+g_{3}a_{L}^{2}(0)a_{A}^{\dagger}(0)\\
 & + & g_{4}a_{V}^{\dagger2}(0)a_{A}(0)+g_{5}a_{S}(0)a_{V}(0)a_{V}^{\dagger}(0)\\
 & + & g_{6}a_{S}(0)a_{L}(0)a_{L}^{\dagger}(0),\\
a_{V}(t) & = & h_{1}a_{V}(0)+h_{2}a_{L}(0)a_{S}^{\dagger}(0)+h_{3}a_{L}^{\dagger}(0)a_{A}(0)\\
 & + & h_{4}a_{S}^{\dagger}(0)a_{V}^{\dagger}(0)a_{A}(0)+h_{5}a_{V}(0)a_{L}(0)a_{L}^{\dagger}(0)\\
 & + & h_{6}a_{V}(0)a_{S}(0)a_{S}^{\dagger}(0)+h_{7}a_{V}(0)a_{A}^{\dagger}(0)a_{A}(0)\\
 & + & h_{8}a_{V}(0)a_{L}^{\dagger}(0)a_{L}(0),\\
a_{A}(t) & = & l_{1}a_{A}(0)+l_{2}a_{L}(0)a_{V}(0)+l_{3}a_{L}^{2}(0)a_{S}^{\dagger}(0)\\
 & + & l_{4}a_{S}(0)a_{V}^{2}(0)+l_{5}a_{V}^{\dagger}(0)a_{V}(0)a_{A}(0)\\
 & + & l_{6}a_{L}(0)a_{L}^{\dagger}(0)a_{A}(0).
\end{array}\label{eq:SM-soln}
\end{equation}
The functional forms of the coefficients in the Sen-Mandal perturbative
solution $f_{i},\,g_{i},\,h_{i},$ and $l_{i}$ is reported in Appendix A for the sake of completeness. Here, we use
the Sen-Mandal solution (\ref{eq:SM-soln}) to write the normal-ordered characteristic
function $C_{N}\equiv C_{N}\left(\beta_{L},\beta_{S},\beta_{V},\beta_{A}\right)$ to describe the Raman process \cite{perina1991quantum}
as 
\begin{equation}
\begin{array}{lcl}
C_{N} & = & \left\langle \exp\left\{ \underset{j=L,S,A,V}{\sum}\left[-B_{j}\left(t\right)\left|\beta_{j}\right|^{2}\right.\right.\right.\\
 & + & \left.\left(\frac{1}{2}C_{j}^{*}\left(t\right)\beta_{j}^{2}+\textrm{c.c.}\right)+\beta_{j}\xi_{j}^{*}\left(t\right)-\beta_{j}^{*}\xi_{j}\left(t\right)\right]\\
 & + & \left.\left.\underset{j<k}{\sum}\left(D_{jk}\left(t\right)\beta_{j}^{*}\beta_{k}^{*}+\overline{D}_{jk}\left(t\right)\beta_{j}\beta_{k}^{*}+\textrm{c.c.}\right)\right\} \right\rangle ,
\end{array}\label{eq:charF}
\end{equation}
where c.c. stands for the complex conjugate. We assume the set $\left\{ L,S,A,V\right\} $
ordered, and the terms $B_{j},\,C_{j},\,D_{jk},$ and $\overline{D}_{jk}$
correspond to quantum noise fluctuations \cite{perina1991quantum}. In the present case, the
obtained quantum noise fluctuation terms, considering all four modes
coherent, are as follows 
\begin{equation}
\begin{array}{lcl}
B_{L}\left(t\right) & = & \left|f_{3}\right|^{2}\left|\xi_{A}\right|^{2},\\
B_{S}\left(t\right) & = & \left|g_{2}\right|^{2}\left|\xi_{L}\right|^{2},\\
B_{V}\left(t\right) & = & \left|h_{2}\right|^{2}\left|\xi_{L}\right|^{2}+\left|h_{3}\right|^{2}\left|\xi_{A}\right|^{2},\\
C_{L}\left(t\right) & = & \left(f_{2}f_{3}+f_{1}f_{4}\right)\xi_{S}\xi_{A},\\
C_{V}\left(t\right) & = & \left(h_{2}h_{3}+h_{1}h_{4}\right)\xi_{S}^{*}\xi_{A},\\
D_{LS}\left(t\right) & = & \left(f_{1}g_{6}+f_{2}g_{2}\right)\xi_{L}\xi_{S},\\
D_{LV}\left(t\right) & = & f_{1}h_{3}\xi_{A}+\left(f_{1}h_{5}+f_{1}h_{8}+f_{2}h_{2}\right)\xi_{L}\xi_{V},\\
D_{LA}\left(t\right) & = & f_{1}l_{6}\xi_{L}\xi_{A},\\
D_{SV}\left(t\right) & = & g_{1}h_{2}\xi_{L}+g_{1}h_{6}\xi_{S}\xi_{V}+\left(g_{1}h_{4}+g_{2}h_{3}\right)\xi_{V}^{*}\xi_{A},\\
D_{SA}\left(t\right) & = & g_{1}l_{3}\xi_{L}^{2},\\
D_{VA}\left(t\right) & = & h_{1}l_{5}\xi_{V}\xi_{A},\\
\overline{D}_{LS}\left(t\right) & = & f_{3}^{*}g_{2}\xi_{L}\xi_{A}^{*}.
\end{array}\label{eq:Terms-in-CF}
\end{equation}
The rest of the terms in Eq. (\ref{eq:charF}) are zero. The solution (\ref{eq:SM-soln})
used to write the characteristic function is already mentioned to be
more general than the short-time solution, and one can verify that the quantum noise terms can be reduced to the corresponding short-time counterparts in the limiting case \cite{pathak2013nonclassicality,perina1991quantum}.
For example, $B_{L}\left(t\right)$ reported
in Eq. (\ref{eq:Terms-in-CF}) can be written as $\left|\chi\right|^{2}t^{2}\left|\xi_{A}\right|^{2}$
up to quadratic terms in $t$, which is same as reported in \cite{pathak2013nonclassicality,perina1991quantum} written in the interaction picture.
It is noteworthy that the perturbative solution (in Eq. (\ref{eq:SM-soln}) and Appendix A) inherently contains
$\Delta\omega_{1}=(\omega_{S}+\omega_{V}-\omega_{L})$ and $\Delta\omega_{2}=(\omega_{L}+\omega_{V}-\omega_{A})$,
which can be defined as frequency detuning in Stokes and anti-Stokes
generation processes, respectively. Thus, in what follows, without loss of generality,
we will consider phase matching conditions to focus only on the
effect of frequency detuning on nonclassical properties.

Even if the Sen-Mandal finite-time solution up to the second order in the interaction
constants leads directly to Gaussian statistics for Raman scattering, we can follow
this matter also from the point of view of its relation to the short-time approximation.
In the finite-time approximation, we substitute time $t$ by $\frac{t[\exp(i\Delta\omega t)-1]}{\Delta\omega t} =\frac{t[\cos (\Delta\omega t) -1 +i \sin(\Delta\omega t)]}{\Delta\omega t} \approx it \textrm{sinc}(\Delta\omega t)$,
as the real part of this expression is small and negligible with increasing frequency detuning. Hence, 
the short time $t$ can be increased up to unity over the coupling constant appropriate for the 
finite-time solution because it is cut by the sinc function. Thus, the finite-time solution describes
frequency transient effects and for frequency tuning the finite-time solution reduces to the short-time
solution. In general, terms in (2) involving $f_4$ and $h_4$ occurring additionally to the short-time terms \cite{perina1991quantum}
are negligible.
In what
follows, we will discuss the nonclassicality in photon and phonon
modes considering all the modes initially coherent. 

\section{Nonclassicality considering phonon mode coherent \label{sec:coh}}

We can use the obtained characteristic function (\ref{eq:charF})
to observe the nonclassicality in all the photon and phonon modes in Raman
process. Here, we use a set of nonclassicality criteria to discuss
the possibility of observing corresponding feature in the process
of our interest. 

\subsection{Intermodal entanglement}

The condition for entanglement can be written as an inequality involving
the terms of the characteristic function defined in Eq. (\ref{eq:Terms-in-CF})
as \cite{perina2011joint}

\begin{equation}
\begin{array}{lcl}
\left(K_{ij}\right)_{\pm} & = & \left(B_{i}\pm\left|C_{i}\right|\right)\left(B_{j}\pm\left|C_{j}\right|\right)-\left(\left|D_{ij}\right|\mp\left|\overline{D}_{ij}\right|\right)^{2}<0.\end{array}\label{eq:cond-ent}
\end{equation}
Specifically, the negative value of even one of these two criteria
is the signature of intermodal entanglement in mode $i$ and $j$.

From the present solution, we have obtained parameter $\left(K_{ij}\right)_{\pm}$
for different combinations, and obtained 

\begin{equation}
\begin{array}{lcl}
\left(K_{LV}\right)_{+} & = & \left(K_{LV}\right)_{-}=-\left|h_{3}\right|^{2}I_{A}<0\end{array}\label{eq:ent-LV}
\end{equation}
and
\begin{equation}
\begin{array}{lcl}
\left(K_{SV}\right)_{+} & = & \left(K_{SV}\right)_{-}=-\left|h_{2}\right|^{2}I_{L}<0,\end{array}\label{eq:ent-SV}
\end{equation}
while the rest of the combinations yield value zero. Here, we have written
$I_{i}=\left|\xi_{i}\right|^{2}$ as the intensity of $i$th mode.
One can notice that the values in the right-hand side for $\left(K_{LV}\right)_{\pm}$
and $\left(K_{SV}\right)_{\pm}$ are always negative. Thus, showing
that the phonon mode is always entangled with both pump and Stokes
mode. While in the domain of the validity of the present solution,
we could not establish the presence of entanglement in the rest of
the combinations. Though the present results do not discard any such
possibility of observing entanglement due to use of a perturbative
solution here. Also note that the Stokes-phonon entanglement can be
observed in spontaneous process as well, while pump-phonon entanglement
can only be observed under partial spontaneous (i.e., when
$I_{A}\neq0$ initially) or stimulated conditions.

We have already mentioned that different terms in the characteristic
function can give corresponding short-time solution terms in the
limiting case. Similarly, the nonclassicality witnesses studied in
Ref. \cite{pathak2013nonclassicality} can also be obtained as a special
case of the expressions obtained here. As a particular example, $\left(K_{LV}\right)_{\pm}=-\left|h_{3}\right|^{2}I_{A}\approx-\left|\chi\right|^{2}t^{2}I_{A}$
for smaller $t$ \cite{pathak2013nonclassicality}. All the results
reported in \cite{pathak2013nonclassicality} can be obtained in the
limiting case of the present expressions, therefore we will not discuss
this point further. 

\subsection{Sub-shot noise}

Another two-mode nonclassical feature is sub-shot noise. The condition
for sub-shot noise can be written as \cite{pathak2013nonclassicality}

\begin{equation}
\begin{array}{lcl}
C_{ij} & = & B_{i}^{2}+B_{j}^{2}+\left|C_{i}\right|^{2}+\left|C_{j}\right|^{2}-2\left|D_{ij}\right|^{2}-2\left|\overline{D}_{ij}\right|^{2}<0.\end{array}\label{eq:cond-sub-shot}
\end{equation}
Interestingly, in the present case, we have obtained using Eq. (\ref{eq:Terms-in-CF})
in Eq. (\ref{eq:cond-sub-shot}) that $C_{ij}=2\left(K_{ij}\right)_{\pm}\,\forall i,j$.
Thus, we have shown that sub-shot noise can be observed in all the
cases when intermodal entanglement is present. Thus, pump-phonon and
Stokes-phonon modes have sub-shot noise.

\subsection{Single-mode and intermodal squeezing}

\begin{figure}
\begin{centering}
\includegraphics[scale=0.5]{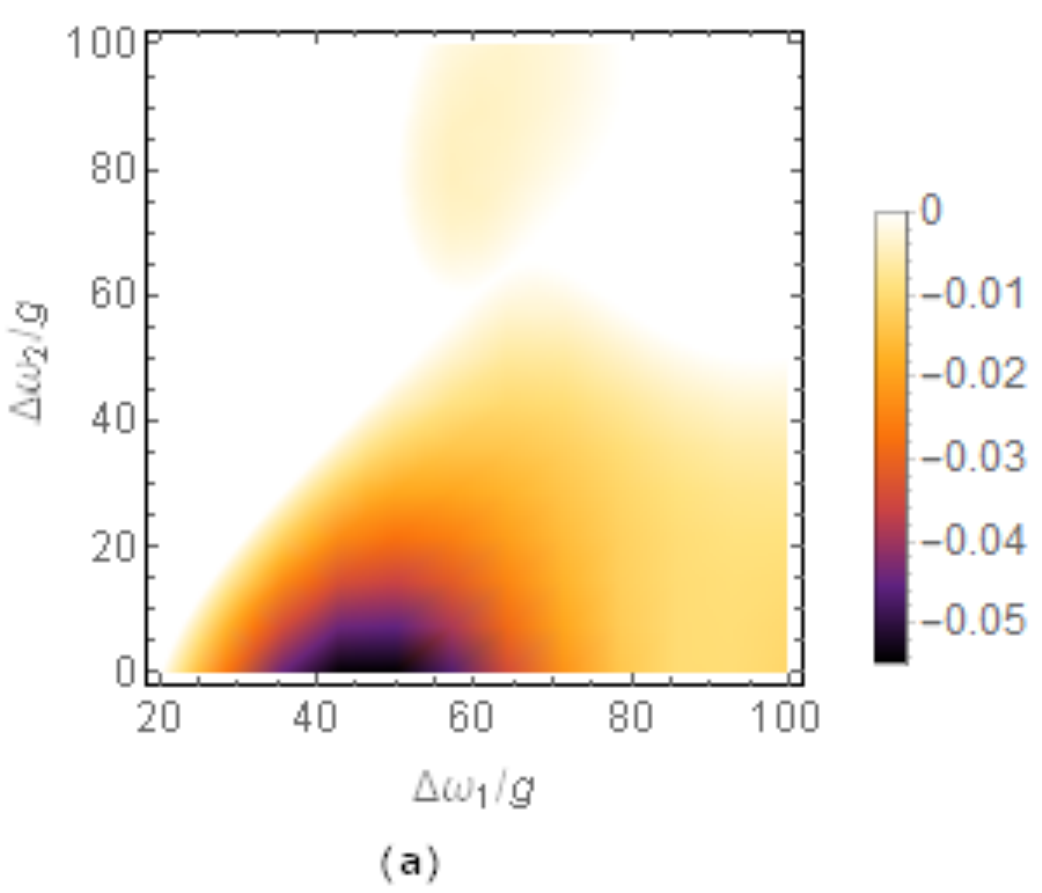} \includegraphics[scale=0.5]{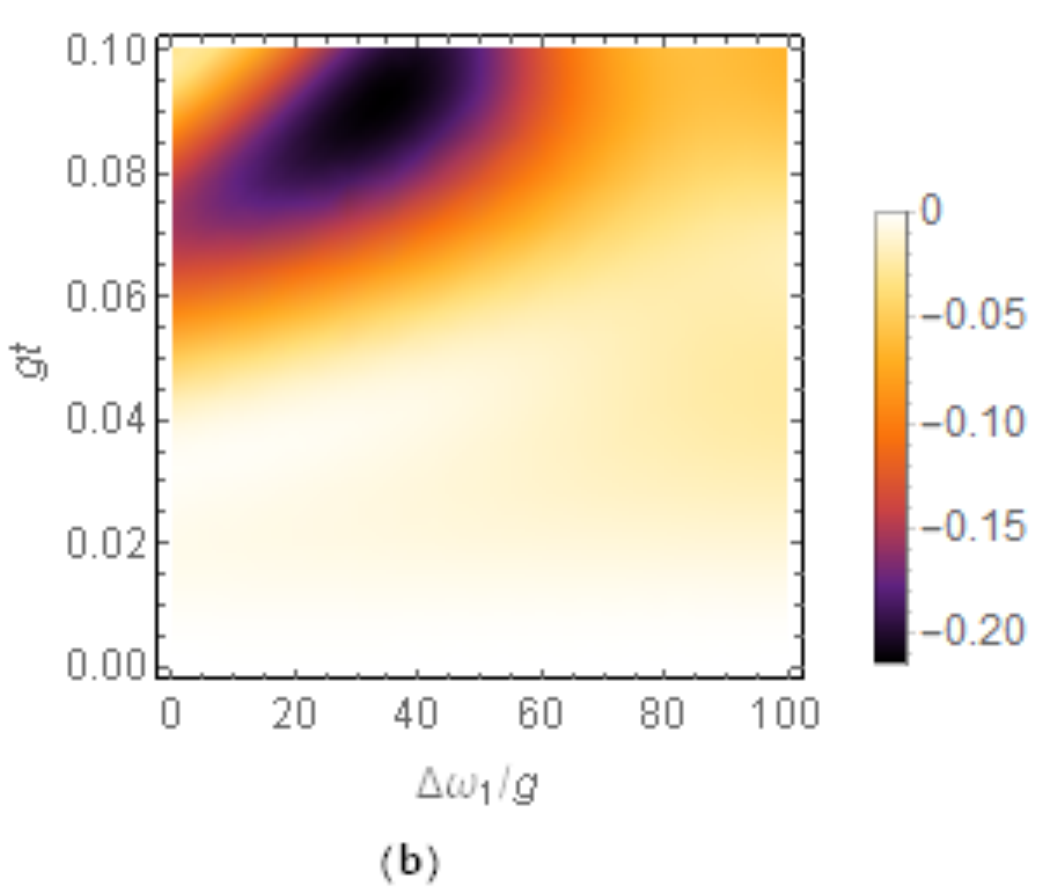}
\par\end{centering}
\begin{centering}
\includegraphics[scale=0.5]{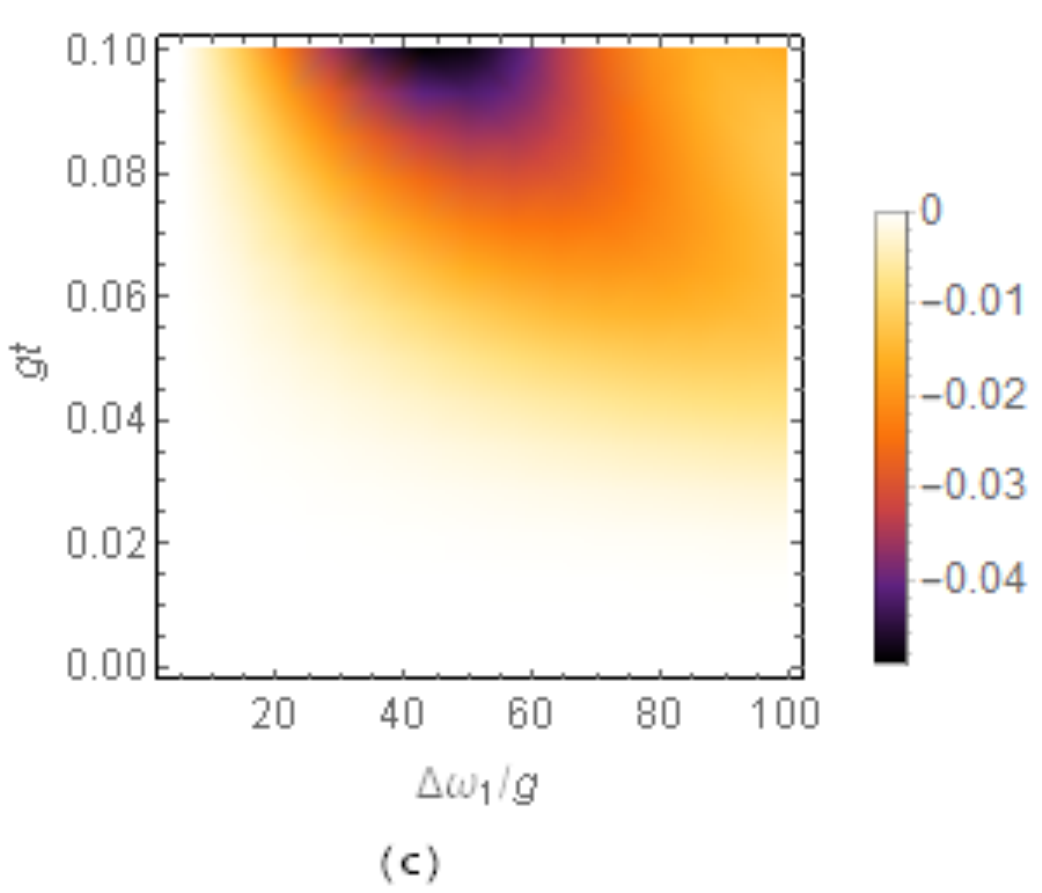} \includegraphics[scale=0.5]{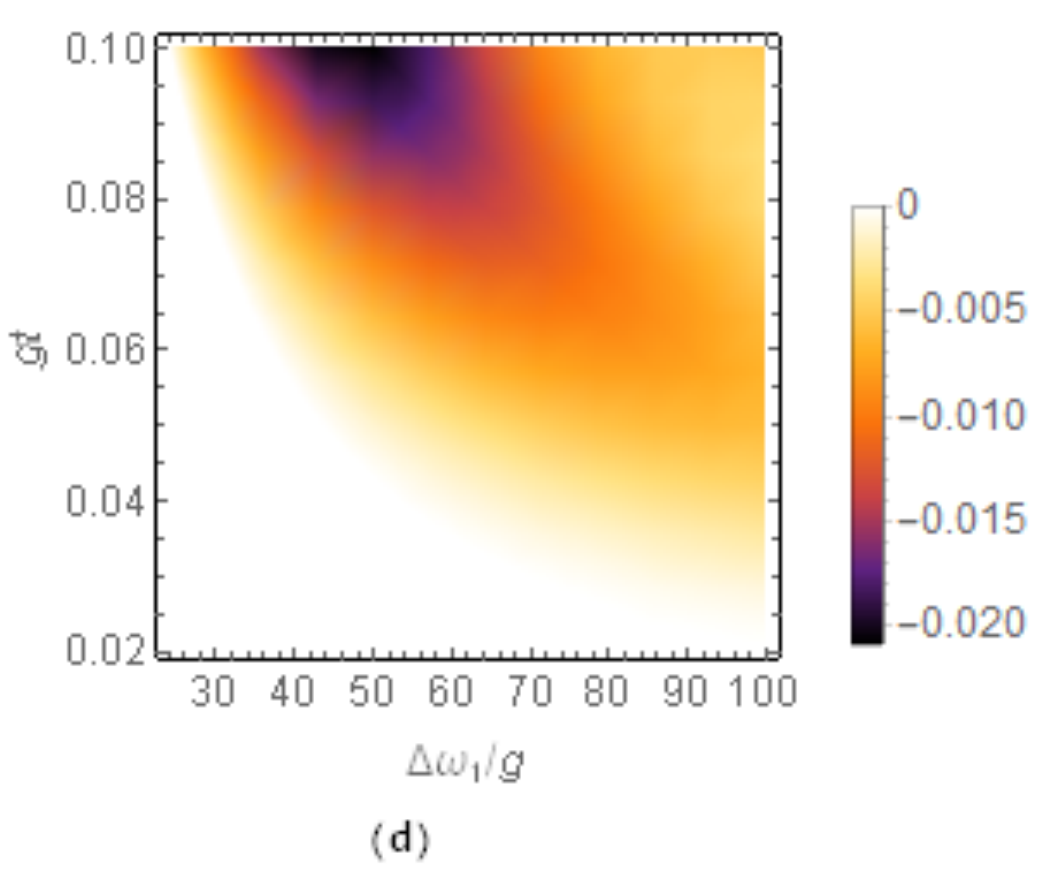}
\par\end{centering}
\caption{\label{fig:sq}(Color online) (a) The negative values of $\lambda_{V}-1$
are shown as function of frequency detuning. Similarly, the time evolution
of the negative values of $\lambda_{LS}-1$, $\lambda_{SA}-1$, and
$\lambda_{VA}-1$ are shown in (b), (c), and (d), respectively. To
obtain the plots, we have assumed $I_{L}=10,\,I_{A}=1,\,I_{S}=9,\,I_{V}=0.01,$ and
$\chi=g$. Wherever needed, we have chosen rescaled time $gt=0.1$
and frequency detuning $\Delta\omega_{2}=10g$. All the quantities shown here and in the rest of the paper are dimensionless.}
\end{figure}

Inspired by the applications of single- and two-mode squeezing \cite{hillery2000quantum,aasi2013enhanced,grote2013first},
we will further study squeezing in these cases. The criteria of single-mode and intermodal squeezing are \cite{perina1991quantum}

\begin{equation}
\begin{array}{lcl}
\lambda_{i} & =1+ & 2\left(B_{i}-\left|C_{i}\right|\right)<1\end{array}\label{eq:cond-sq}
\end{equation}
and
\begin{equation}
\begin{array}{lcl}
\lambda_{ij} & =1+ & B_{i}+B_{j}-2\textrm{Re}\left|\overline{D}_{ij}\right|-\left|C_{i}+C_{j}+2D_{ij}\right|<1,\end{array}\label{eq:cond-int-sq}
\end{equation}
respectively. Using Eq. (\ref{eq:Terms-in-CF}) in Eq. (\ref{eq:cond-sq}),
we obtained the following expressions for squeezing witnesses corresponding to
the pump and phonon modes
\begin{equation}
\begin{array}{lcl}
\lambda_{L} & = & 1+2\left|f_{3}\right|^{2}I_{A}-2\left|f_{2}f_{3}+f_{1}f_{4}\right|\left|\xi_{S}\right|\left|\xi_{A}\right|,\\
\lambda_{V} & = & 1+2\left|h_{2}\right|^{2}I_{L}+2\left|h_{3}\right|^{2}I_{A}-2\left|h_{2}h_{3}+h_{1}h_{4}\right|\left|\xi_{S}\right|\left|\xi_{A}\right|,
\end{array}\label{eq:sq}
\end{equation}
while the rest of $\lambda_{i}\ge1$. Therefore, one can clearly conclude
that Stokes and anti-Stokes modes do not show any signature of squeezing.
From the obtained expressions for squeezing (\ref{eq:sq}), we obtained
that $\lambda_{L}<1$ which is consistent with the nature reported
in corresponding short-time case \cite{pathak2013nonclassicality}.
However, for the Stokes modes, short-time solution failed to detect
squeezing whereas analysis of the obtained result (\ref{eq:sq}) revealed
that $\lambda_{V}<1$ for specific values of frequency detuning. From
Fig. \ref{fig:sq} (a), one can clearly see that with a proper choice
of frequency detuning in Stokes generation process, squeezing can
be observed in the phonon mode as well. This shows the advantage of
using a more general solution than short-time solution as it has successfully
detected squeezing which was not observed previously \cite{pathak2013nonclassicality}. This also establishes the frequency detuning parameter as a control to enhance/induce nonclassicality in the output of scattering process.
The observed nonclassicality disappears in both spontaneous and partial
spontaneous cases.

Similarly, using Eq. (\ref{eq:Terms-in-CF}) in Eq. (\ref{eq:cond-int-sq}),
the expressions for intermodal squeezing in all the cases can be obtained.
Here, we are not giving the expressions and would like to emphasize
the relevance of frequency detuning and more general nature of solution.
Specifically, intermodal squeezing in pump-phonon, pump-anti-Stokes,
and Stokes-phonon modes were established through short-time solution
in the past \cite{pathak2013nonclassicality}. Here, we have not only
observed these nonclassical features, we have also established the
presence of intermodal squeezing in pump-Stokes, Stokes-anti-Stokes,
and phonon-anti-Stokes modes, which can be observed to depend upon
the frequency detuning in Stokes generation (cf. Fig. \ref{fig:sq}
(b)-(d)). In case of squeezing observed in the pump-Stokes mode (in Fig. \ref{fig:sq}
(b)), it
can also be attributed to the more general nature of solution, which
contains the negative terms dominating for longer time evolution.

\subsection{Wave variances}

Using the Gaussian behavior of the obtained normal characteristic
function (\ref{eq:charF}), we can also compute the fluctuation quantities
in terms of variances $\left\langle \left(\Delta W_{i}\right)^{2}\right\rangle $
and $\left\langle \Delta W_{i}\Delta W_{j}\right\rangle $ as \cite{perina1991quantum}
\begin{equation}
\begin{array}{lcl}
\left\langle \left(\Delta W_{i}\right)^{2}\right\rangle &=&\left\langle B_{i}^{2}+\left|C_{i}\right|^{2}+2B_{i}\left|\xi_{i}\left(t\right)\right|^{2}\right.\\&+&\left.\left\{ C_{i}\xi_{i}^{*2}\left(t\right)+\textrm{c.c.}\right\} \right\rangle  \label{eq:variance}
\end{array}
\end{equation}
and 
\begin{equation}
\begin{array}{lcl}
\left\langle \Delta W_{i}\Delta W_{j}\right\rangle &=&\left\langle \left|D_{ij}\right|^{2}-\left|\overline{D}_{ij}\right|^{2}+ \left\{ D_{ij}\xi_{i}^{*}\left(t\right)\xi_{j}^{*}\left(t\right)\right.\right.\\
&- &\left.\left.\overline{D}_{ij}\xi_{i}\left(t\right)\xi_{j}^{*}\left(t\right)+\textrm{c.c.}\right\} \right\rangle.\label{eq:2vari}\end{array}
\end{equation}
The single-mode fluctuation quantities using Eq. (\ref{eq:Terms-in-CF})
in Eq. (\ref{eq:variance}) lead to
\begin{equation}
\begin{array}{lcl}
\left\langle \left(\Delta W_{L}\right)^{2}\right\rangle  & = & 2\left|f_{3}\right|^{2}I_{A}I_{L}+\left\{ \left(f_{2}f_{3}+f_{1}f_{4}\right)\xi_{S}\xi_{A}\xi_{L}^{*2}+\textrm{c.c.}\right\}\\
\left\langle \left(\Delta W_{V}\right)^{2}\right\rangle  & = & 2\left(\left|h_{2}\right|^{2}I_{L}+\left|h_{3}\right|^{2}I_{A}\right)I_{V}\\
 & + &\left\{ \left(h_{2}h_{3}+h_{1}h_{4}\right)\xi_{S}^{*}\xi_{A}\xi_{V}^{*2}+\textrm{c.c.}\right\} 
\end{array}\label{eq:vari-sin}
\end{equation}
for pump and phonon mode, respectively. The pump mode shows antibunching (i.e.,
has negative values of fluctuation) for both zero (shown in \cite{perina1991quantum})
and non-zero (present case) values of frequency detuning. However,
Stokes and anti-Stokes modes fail to show antibunching in both cases.
While the phonons can be shown to exhibit antibunching for large
values of detuning in the Stokes generation. Nonclassicality in both these cases can be observed to disappear in spontaneous case.

These variances further allow us to calculate nonclassical sum- or
difference-variances defined as \cite{perina1991quantum}
\begin{equation}
\left\langle \left(\Delta W_{ij}\right)^{2}\right\rangle _{\pm}=\left\langle \left(\Delta W_{i}\right)^{2}\right\rangle +\left\langle \left(\Delta W_{j}\right)^{2}\right\rangle \pm2\left\langle \Delta W_{i}\Delta W_{j}\right\rangle <0.\label{eq:SDvari}
\end{equation}
In the present case, we have computed the sum and difference variances
using Eq. (\ref{eq:Terms-in-CF}) in Eqs. (\ref{eq:variance}), (\ref{eq:2vari}), and (\ref{eq:SDvari}),
and observed nonclassicalicality reflected through this witness in
all combinations. Note that the previous studies were successful in
detecting the nonclassicality reflected through this criteria only
in some cases \cite{pathak2013nonclassicality}. Specificly, due
to frequency detuning parameters being non-zero in our present solution,
wave variances also become functions of frequency of the corresponding
modes as well. Particularly, the value of the frequency of the phonon
mode is expected to be much smaller compared to that of the photon
modes, and thus leads to fast oscillations in the short-time scale
while revealing corresponding nonclassical behavior. Therefore,
the present study revealed that frequency detuning can play an important
role as it leads to the presence of abundant nonclassical features. 

Nonclassical properties in the Raman and hyper-Raman processes studied using a different set of criteria 
could
not demonstrate most of the results reported here as the effect of frequency detuning was not included there \cite{thapliyal2017nonclassicality}.
In what follows, we will consider an interesting case, i.e., when the phonon
mode is initially chaotic. 

\section{Nonclassicality considering phonon mode chaotic \label{sec:cha}}

We can further obtain the characteristic function considering the
phonon mode chaotic (with average phonon number $\left\langle n_{V}\right\rangle $),
while the rest of the modes are initially in coherent state, to observe
the nonclassicality in the states generated in the off-resonant Raman process. The form of the characteristic function remains unchanged,
i.e., can be defined by Eq. (\ref{eq:charF}), while different terms
in (\ref{eq:charF}) are now defined as
\begin{equation}
\begin{array}{lcl}
B_{L}\left(t\right) & = & \left|f_{2}\right|^{2}\left|\xi_{S}\right|^{2}\left\langle n_{V}\right\rangle +\left|f_{3}\right|^{2}\left|\xi_{A}\right|^{2}\left(\left\langle n_{V}\right\rangle +1\right),\\
B_{S}\left(t\right) & = & \left|g_{2}\right|^{2}\left|\xi_{L}\right|^{2}\left(\left\langle n_{V}\right\rangle +1\right),\\
B_{V}\left(t\right) & = & \left\langle n_{V}\right\rangle +\left|h_{2}\right|^{2}\left(\left|\xi_{L}\right|^{2}+\left\langle n_{V}\right\rangle \left\{ \left|\xi_{L}\right|^{2}-\left|\xi_{S}\right|^{2}\right\} \right)\\
&  +&  \left|h_{3}\right|^{2}\left(\left|\xi_{A}\right|^{2}+\left\langle n_{V}\right\rangle \left\{ \left|\xi_{A}\right|^{2}-\left|\xi_{L}\right|^{2}\right\} \right),\\
B_{A}\left(t\right) & = & \left|l_{2}\right|^{2}\left|\xi_{L}\right|^{2}\left\langle n_{V}\right\rangle ,\\
C_{L}\left(t\right) & = & \left(f_{2}f_{3}\left\{ 2\left\langle n_{V}\right\rangle +1\right\} +f_{1}f_{4}\right)\xi_{S}\xi_{A},\\
C_{V}\left(t\right) & = & \left(h_{2}h_{3}+h_{1}h_{4}\left\{ 2\left\langle n_{V}\right\rangle +1\right\} \right)\xi_{S}^{*}\xi_{A},\\
D_{LS}\left(t\right) & = & \left(f_{1}g_{6}+f_{2}g_{2}\left\{ \left\langle n_{V}\right\rangle +1\right\} \right)\xi_{L}\xi_{S},\\
D_{LV}\left(t\right) & = & f_{1}h_{3}\left(\left\langle n_{V}\right\rangle +1\right)\xi_{A},\\
D_{LA}\left(t\right) & = & \left(f_{1}l_{6}+f_{3}l_{2}\left\langle n_{V}\right\rangle \right)\xi_{L}\xi_{A},\\
D_{SV}\left(t\right) & = & g_{1}h_{2}\left(\left\langle n_{V}\right\rangle +1\right)\xi_{L},\\
D_{SA}\left(t\right) & = & \left(g_{1}l_{3}+g_{2}h_{2}\left\langle n_{V}\right\rangle \right)\xi_{L}^{2},\\
\overline{D}_{LS}\left(t\right) & = & f_{3}^{*}g_{2}\left(\left\langle n_{V}\right\rangle +1\right)\xi_{L}\xi_{A}^{*},\\
\overline{D}_{LV}\left(t\right) & = & f_{2}^{*}h_{1}\left\langle n_{V}\right\rangle \xi_{S}^{*},\\
\overline{D}_{LA}\left(t\right) & = & f_{2}^{*}l_{2}\left\langle n_{V}\right\rangle \xi_{L}\xi_{S}^{*},\\
\overline{D}_{VA}\left(t\right) & = & h_{1}l_{2}^{*}\left\langle n_{V}\right\rangle \xi_{L}.
\end{array}\label{eq:Terms-in-CF-Chaotic}
\end{equation}
The rest of the terms in Eq. (\ref{eq:charF}) are zero for the present
case. The functional forms of the coefficients used here are given in Eqs. (\ref{eq:solutions of f})-(\ref{eq:eq:solutions of l})
in Appendix A. We further use the obtained characteristic function
defined by terms (\ref{eq:Terms-in-CF-Chaotic}) considering phonon
mode chaotic to observe the nonclassicality in the states generated
in the off-resonant Raman process with specific interest in the frequency detuning
parameter. It is worth mentioning here that the characteristic functions
obtained considering phonon mode coherent and chaotic match exactly
if the phonon mode is initially in vacuum state, i.e., $\xi_{V}=\left\langle n_{V}\right\rangle =0$.
This fact serves as a consistency check for the obtained solutions and
will be used in the last section to study joint-photon number and
integrated intensity distributions in this case. The obtained characteristic
function (\ref{eq:Terms-in-CF-Chaotic}) can further be reduced to
corresponding short-time case \cite{pathak2013nonclassicality}, where
$B_{V}\left(t\right)\approx\left\langle n_{V}\right\rangle $.

\subsection{Intermodal entanglement}

Using the condition for entanglement (\ref{eq:cond-ent}) and the
terms of the characteristic function defined in Eq. (\ref{eq:Terms-in-CF-Chaotic}),
we have obtained the parameter $\left(K_{ij}\right)_{\pm}$ for different
combinations, and obtained 

\begin{equation}
\begin{array}{lcl}
\left(K_{LV}\right)_{\pm} & = & -\left|h_{3}\right|^{2}I_{A}\left(\left\langle n_{v}\right\rangle +1\right)\mp\left(\left|f_{2}\right|\left|f_{3}\right|-\left|f_{1}\right|\left|f_{4}\right|\right)\\
&\times&\left\langle n_{v}\right\rangle \left|\xi_{S}\right|\left|\xi_{A}\right|\end{array}\label{eq:ent-LV-ch}
\end{equation}
and
\begin{equation}
\begin{array}{lcl}
\left(K_{SV}\right)_{\pm} & = & -\left|h_{2}\right|^{2}I_{L}\left(\left\langle n_{v}\right\rangle +1\right)<0,\end{array}\label{eq:ent-SV-ch}
\end{equation}
while the rest of the combinations yield value zero. One can clearly observe
that $\left(K_{SV}\right)_{\pm}$ are always negative, thus showing
that Stokes-photon mode is always entangled.
It is noteworthy that we have observed the same behavior considering
phonon mode coherent earlier (in Eq. (\ref{eq:ent-SV})), here the
value of the entanglement witness is further enhanced due to factor
$\left(\left\langle n_{v}\right\rangle +1\right)$, which is expected
to help in the experimental verification of the present result. Unlike
this, $\left(K_{LV}\right)_{-}$ fail to show entanglement, but $\left(K_{LV}\right)_{+}$
is always negative and thus establishes the inseparability of pump
and phonon modes. Hence, the present study revealed that the phonon
mode is always entangled with both pump and Stokes modes. Further,
in case of spontaneous process, only Stokes-phonon entanglement could
be observed, and the results are non-conclusive in the rest of the cases.
The present result establishes
that entanglement (in Stokes-phonon and pump-phonon modes) is  neither restricted to
short-time regime nor frequency matching conditions and thus eliminates these restrictions. 

\subsection{Sub-shot noise}

Using the condition for sub-shot noise (\ref{eq:cond-sub-shot}) with
Eq. (\ref{eq:Terms-in-CF-Chaotic}) we have obtained $C_{ij}\,\forall i,j$,
as 
\begin{equation}
\begin{array}{lcl}
C_{LV} & = & \left\langle n_{v}\right\rangle ^{2}-2\left\{ \left|h_{3}\right|^{2}I_{A}\left(\left\langle n_{v}\right\rangle +1\right)^{2}+\left|f_{2}\right|^{2}\left\langle n_{v}\right\rangle ^{2}I_{S}\right\},\\
C_{SV} & = & \left\langle n_{v}\right\rangle ^{2}-2\left|h_{2}\right|^{2}I_{L}\left(\left\langle n_{v}\right\rangle +1\right)^{2},\\
C_{VA} & = & \left\langle n_{v}\right\rangle ^{2}-2\left|l_{2}\right|^{2}\left\langle n_{v}\right\rangle ^{2}I_{L},
\end{array}\label{eq:sub-shot-ch}
\end{equation}
while the rest of the terms are zero. This shows the presence of sub-shot
noise nonclassical feature even in the absence of intermodal entanglement
(in anti-Stokes-phonon mode) and thus it establishes the relevance of
studying entanglement and sub-shot noise separately. Thus, pump-phonon,
Stokes-phonon, and anti-Stokes-phonon modes exhibit sub-shot noise. Notice
that the presence of nonclassicality can be established in all these combinations
for the partial spontaneous process (i.e., considering nonzero $I_{A}$),
while except pump-phonon mode the rest of the modes show this feature
even in spontaneous case. 

\subsection{Single-mode and intermodal squeezing}

\begin{figure}
\begin{centering}
\includegraphics[scale=0.6]{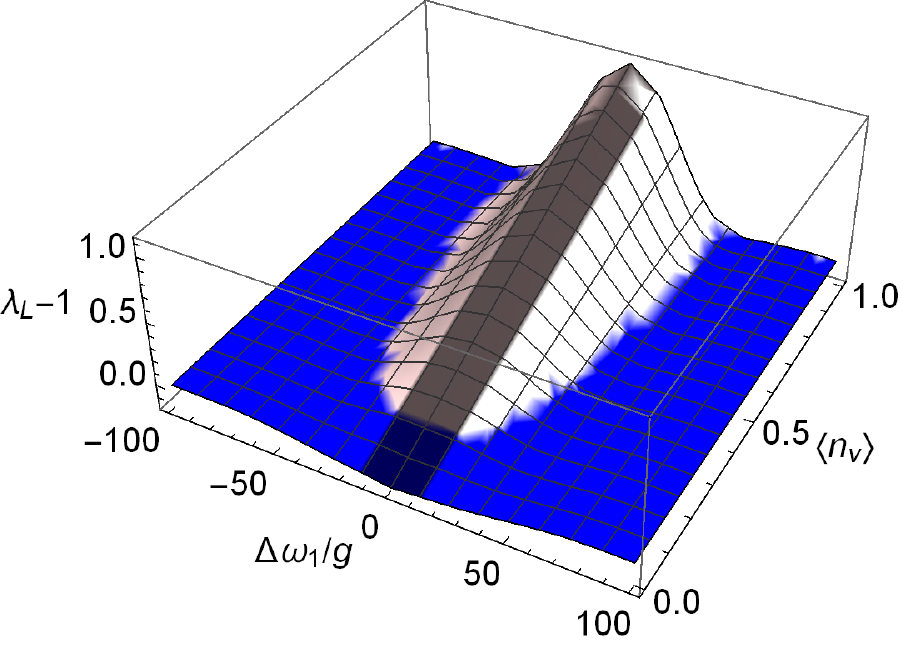}
\par\end{centering}
\caption{\label{fig:sq-ch}(Color online) The negative (positive) values of
$\lambda_{L}-1$ are shown as function of frequency detuning $\Delta\omega_{1}$
and average phonon number by dark blue (light red) colored regions
in the surface plots. To obtain the plot, we have assumed $I_{L}=10,\,I_{A}=1,\,I_{S}=9,$
and $\chi=g$ with the rescaled time $gt=0.1$ and frequency detuning
$\Delta\omega_{2}=10g$. }
\end{figure}

We will further study single-mode and intermodal squeezing with chaotic
phonons. We restrict our discussion here on the nonclassicality observed
in the modes except phonon mode. Using Eq.
(\ref{eq:Terms-in-CF-Chaotic}) in Eq. (\ref{eq:cond-sq}), we obtained
\begin{equation}
\begin{array}{lcl}
\lambda_{L} & = & 1+2\left\{ \left|f_{3}\right|^{2}I_{A}\left(\left\langle n_{v}\right\rangle +1\right)+\left|f_{2}\right|^{2}\left\langle n_{v}\right\rangle I_{S}\right\} \\
&-&2\left|f_{2}f_{3}\left(2\left\langle n_{v}\right\rangle +1\right)+f_{1}f_{4}\right|\left|\xi_{S}\right|\left|\xi_{A}\right|,\end{array}\label{eq:sq-ch}
\end{equation}
while the squeezing parameter is always greater than 1 for Stokes
and anti-Stokes modes and thus do not show any signature of squeezing
in these cases even with chaotic phonon. From the obtained expression (\ref{eq:sq-ch}) for the pump mode, we have observed
that nonclassicality can be obtained (cf. Fig. \ref{fig:sq-ch}). Clearly,
the presence of nonclassicality prefers lower values of average phonon
number and higher values of frequency detuning. 

Similarly, using Eq. (\ref{eq:Terms-in-CF-Chaotic}) in Eq. (\ref{eq:cond-int-sq}),
the expressions for intermodal squeezing are obtained for all the
cases except involving phonon mode and observed that nonclassicality
can be observed. Here, we refrain from further discussion of these
cases. 

\subsection{Wave variances}

The fluctuation quantities using Eqs. (\ref{eq:Terms-in-CF-Chaotic})
in Eq. (\ref{eq:variance}) for chaotic phonon lead to
\begin{equation}
\begin{array}{lcl}
\left\langle \left(\Delta W_{L}\right)^{2}\right\rangle  & = & 2\left\{ \left|f_{3}\right|^{2}I_{A}\left(\left\langle n_{v}\right\rangle +1\right)+\left|f_{2}\right|^{2}\left\langle n_{v}\right\rangle I_{S}\right\} I_{L}\\
&+&\left\{ \left(f_{2}f_{3}\left(2\left\langle n_{v}\right\rangle +1\right)+f_{1}f_{4}\right)\xi_{S}\xi_{A}\xi_{L}^{*2}+\textrm{c.c.}\right\} ,\end{array}\label{eq:vari-sin-1}
\end{equation}
in case of pump mode, while the rest of the modes did not show nonclassicality.
Specifically, in the pump mode lower values of average phonon number
and higher values of frequency detuning are preferable. The correlation
fluctuation can also be computed in this case using Eq. (\ref{eq:variance}).
We have further calculated the sum and difference variances using
Eq. (\ref{eq:SDvari}) and observed that for higher values of frequency
detuning this nonclassical feature can be observed for even large
average phonon numbers.

\section{Joint photon-phonon number and wave distributions \label{sec:joint}}

Nonclassicality in several processes has been studied using joint
photon number and integrated intensity distributions (\cite{perina2011joint,pathak2013nonclassicality}
and references therein). Here, we have performed a similar study for
the off-resonant Raman process assuming phonon vacuum (justified at room temperature),
i.e., $\left\langle n_{v}\right\rangle =0$.

\begin{figure}
\begin{centering}
\includegraphics[scale=0.6]{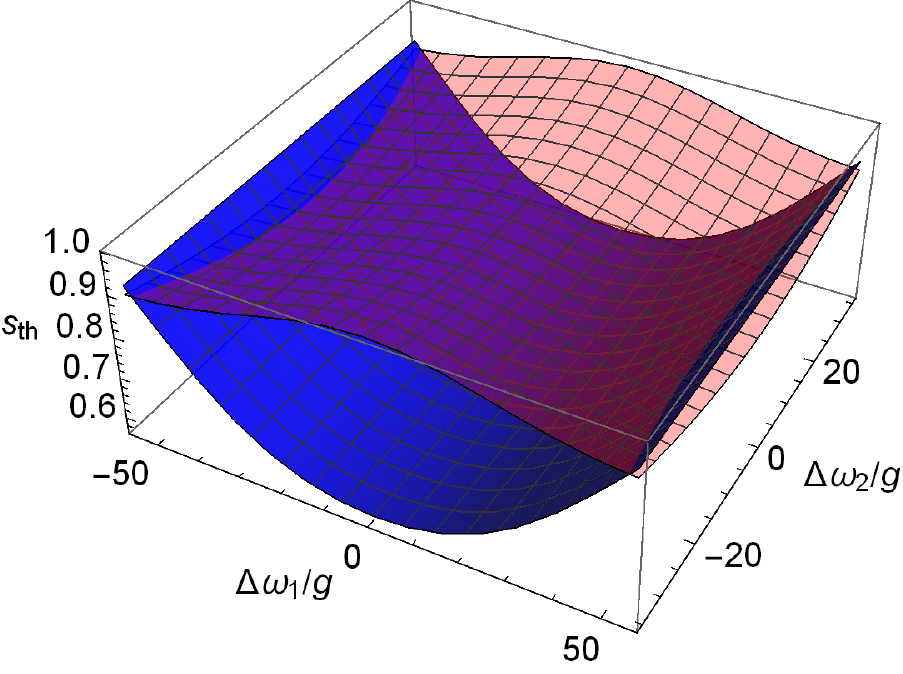}
\par\end{centering}
\caption{\label{fig:sth-det}(Color online) Variations of threshold parameter
for quantum features using integrated intensity distribution for Stokes-phonon
and pump-phonon modes are represented by the blue (at bottom) and red
(on top) surface plots, respectively. To obtain this plot, we have
assumed $I_{L}=10,\,I_{A}=1$ with $\chi=g$ at a dimensionless time
$gt=0.1$. }
\end{figure}

Interestingly, under this condition, for the spontaneous process,
$B_{L}=B_{V}=-K_{LV}$ while the rest of the parameters in characteristic
function (\ref{eq:Terms-in-CF-Chaotic}) for pump and phonon modes
are zero. Using this fact, we can write the joint photon-phonon number
distribution \cite{perina2011joint} as 
\begin{equation}
p\left(n_{S},n_{V}\right)=\frac{\left(B_{S}\right)^{n_{S}}}{\left(1+B_{S}\right)^{1+n_{S}}}\delta_{n_{S},n_{V}},\label{eq:pnsv}
\end{equation}
which clearly indicates pair generation. Corresponding $s$ ordered integrated intensity quasidistribution
\cite{perina2011joint} is obtained as

\begin{equation}
P_{s}\left(W_{S},W_{V}\right)=\frac{1}{\pi B_{Ss}}\exp\left(-\frac{W_{S}+W_{V}}{2B_{Ss}}\right)\frac{\sin\left(\frac{W_{S}-W_{V}}{\sqrt{-K_{SVs}}}\right)}{W_{S}-W_{V}}.\label{eq:pssv}
\end{equation}
Here, $B_{Ss}=B_{S}+\frac{1-s}{2}$ and $K_{SVs}=K_{SV}+\left(1-s\right)B_{s}+\frac{\left(1-s\right)^{2}}{4}$
with $s$ as an ordering parameter. A threshold value of the $s$
parameter can be calculated as $s_{\textrm{th}}$ below (above) which
the obtained quasidistribution behaves like a classical probability
distribution  (it can have negative values and thus shows signature of
quantumness). The threshold value is calculated in this case as $s_{\textrm{th}}=1+2B_{S}-2\sqrt{B_{S}}$,
shown by the blue surface plot in Fig. \ref{fig:sth-det}, where we
can clearly see the value of this parameter to increase (and thus parametric region of nonclassicality to decrease) with increasing frequency
detuning. Therefore, the $s$ ordered integrated intensity distribution
shows the nonclassical behavior for larger range of $s$ if the frequency
matching condition is satisfied. 

In case of the pump-phonon mode, considering the partial spontaneous process,
i.e., $I_{A}>0$ while $I_{S}=0$, we have $B_{L}=B_{V}-B_{S}=-K_{LV}$
while the rest of the parameters for both pump and phonon modes are zero.
With the help of these values, we can write the joint pump-phonon
number distribution as 
\begin{equation}
p\left(n_{L},n_{V}\right)=\frac{n_{V}!}{n_{L}!\left(n_{V}!-n_{L}!\right)}\frac{\left(B_{L}\right)^{n_{L}}\left(B_{V}-B_{L}\right)^{n_{V}-n_{L}}}{\left(1+B_{V}\right)^{1+n_{V}}},\label{eq:pnlv}
\end{equation}
if $n_{V}\ge n_{L}$, and the distribution
is zero for $n_{V}<n_{L}$. We further calculate the threshold
value of $s$ parameter in the present case and obtain $s_{\textrm{th}}=1+2B_{L}+B_{S}-2\sqrt{B_{L}}.$
As shown in Fig. \ref{fig:sth-det}, the value of this parameter does
not vary as rapidly as in the previous case of Stokes-phonon mode.
However, we can observe the effect of frequency detuning which, in
sharp contrast to the previous case, shows an increase in the nonclassical
regime with increasing frequency detuning. In fact, for large frequency
detuning in Stokes generation pump-phonon mode has a lower value of the threshold parameter than corresponding parameter for Stokes-phonon
mode. This shows that the frequency detuning in Stokes generation favors
nonclassicality observed in pump-phonon mode, but suppresses it in
Stokes-phonon mode. Due to very small variation in $s_{\textrm{th}}$
for pump-phonon mode, we have chosen $s=1$ here and obtained Glauber-Sudarshan
integrated intensity quasidistribution for pump-phonon mode as

\begin{equation}
\begin{array}{lcl}
P_{N}\left(W_{L},W_{V}\right)&=&\frac{1}{\pi\sqrt{B_{L}B_{V}}}\exp\left(-\frac{W_{L}}{2B_{L}}-\frac{W_{V}}{2B_{V}}\right)\\
&\times&\frac{\sin\left(\frac{\sqrt{\frac{B_{V}}{B_{L}}}W_{L}-\sqrt{\frac{B_{L}}{B_{V}}}W_{V}}{\sqrt{B_{L}}}\right)}{\sqrt{\frac{B_{V}}{B_{L}}}W_{L}-\sqrt{\frac{B_{L}}{B_{V}}}W_{V}}.\label{eq:pslv}
\end{array}
\end{equation}

\begin{figure}
\begin{centering}
\includegraphics[scale=0.5]{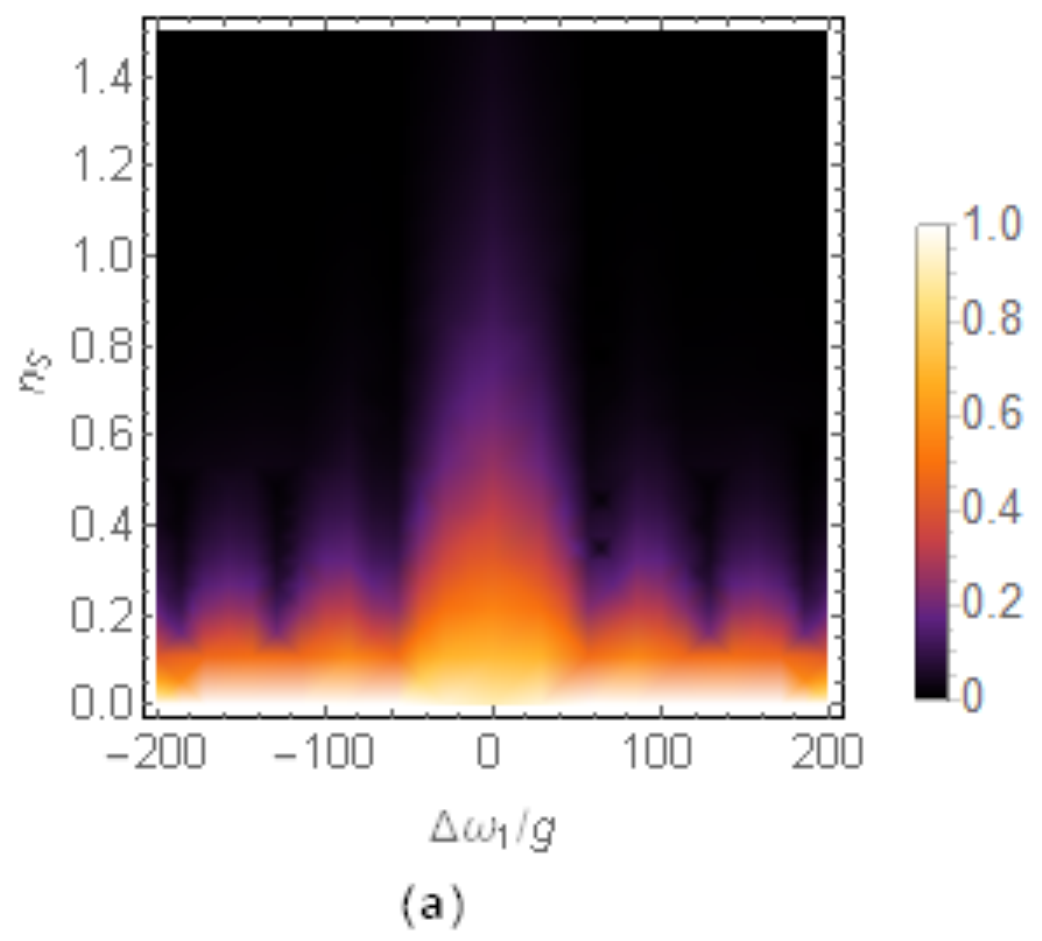} \includegraphics[scale=0.5]{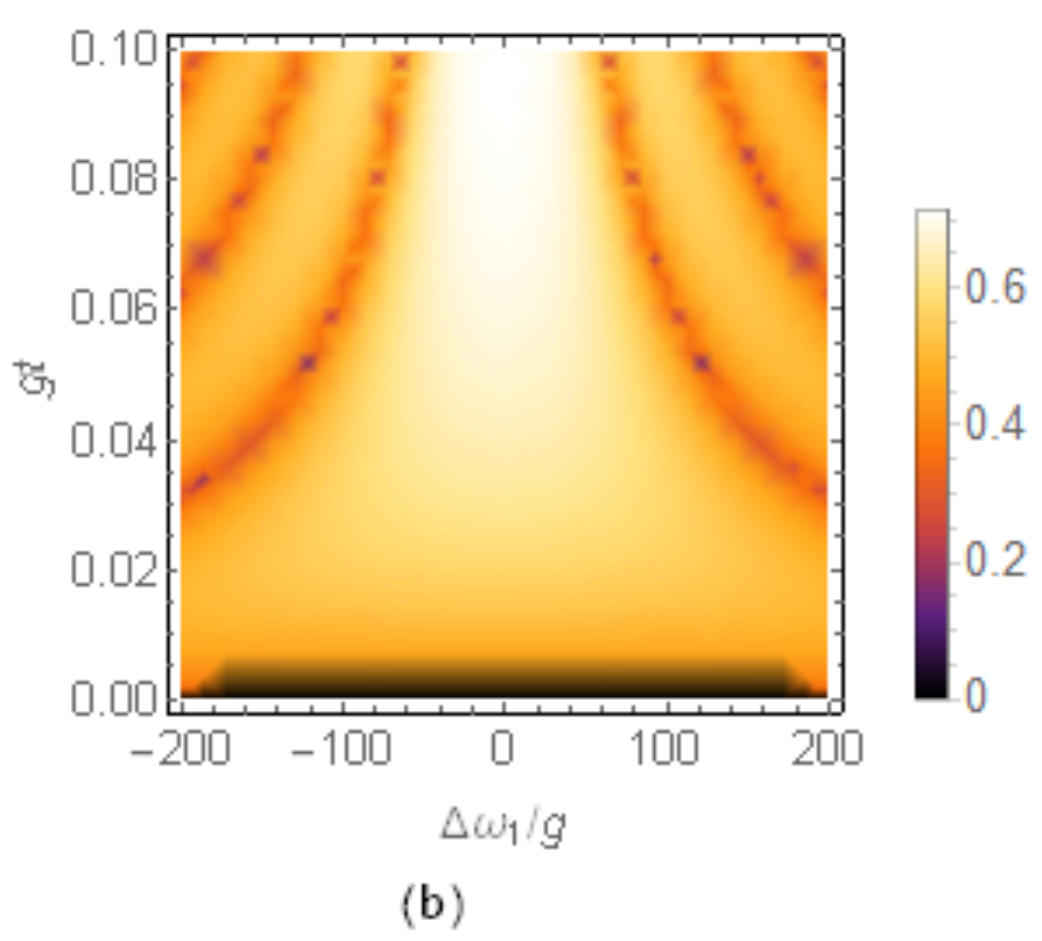}
\par\end{centering}
\begin{centering}
\includegraphics[scale=0.56]{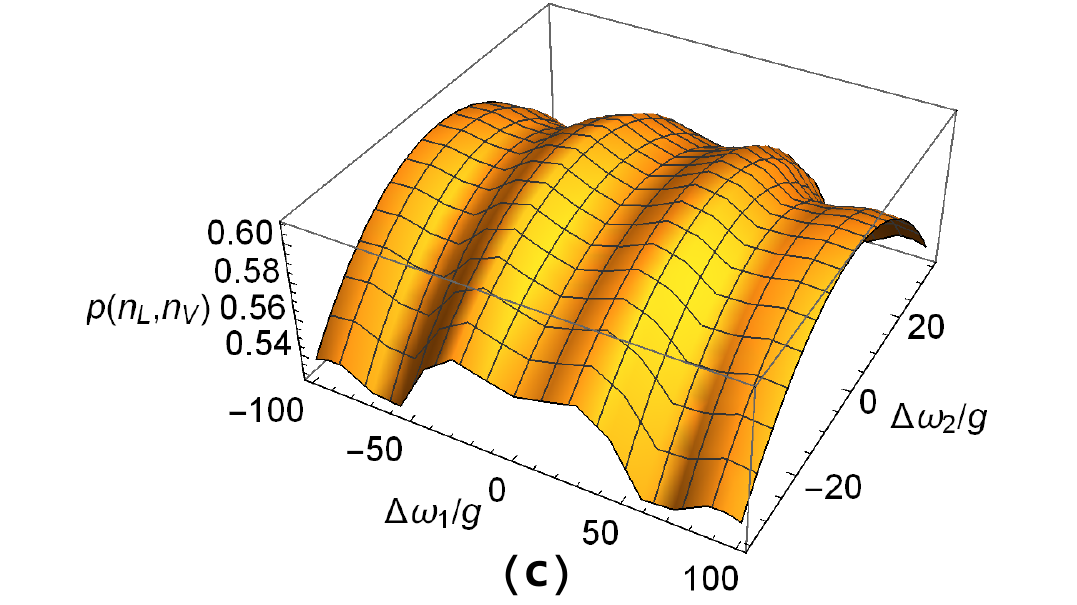} \includegraphics[scale=0.5]{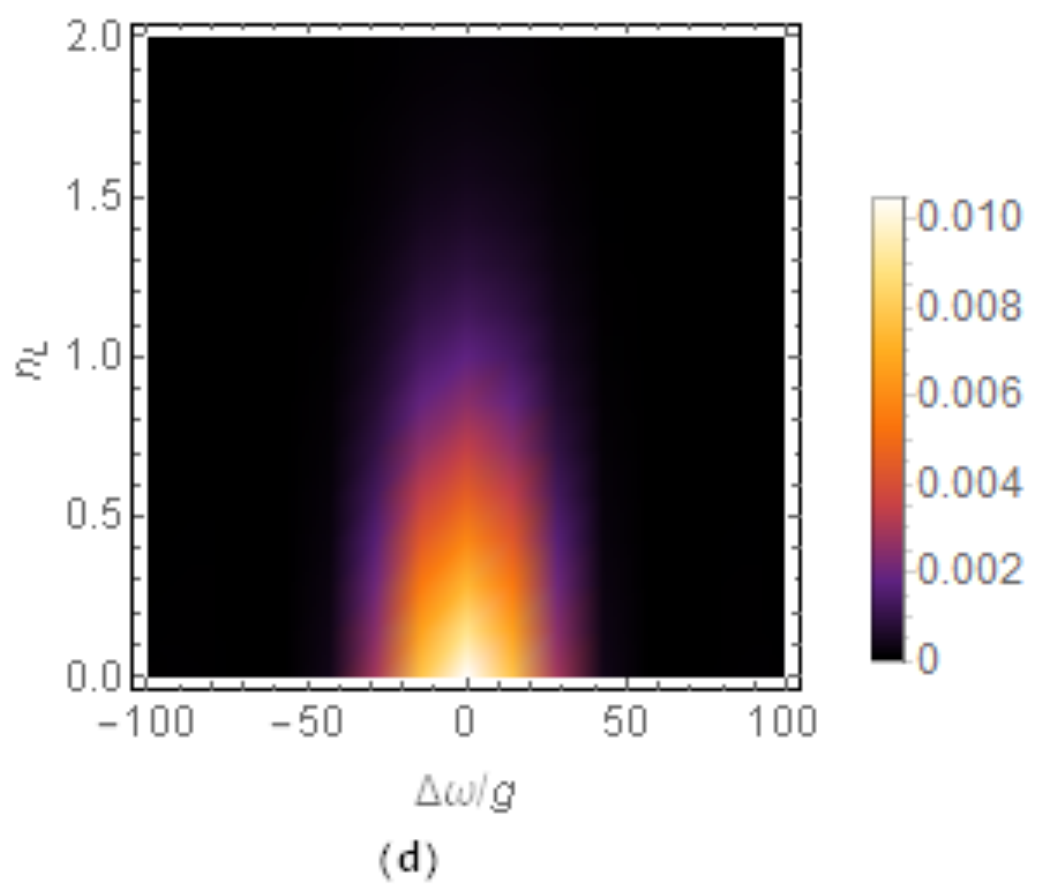}
\par\end{centering}
\caption{\label{fig:joint-det}(Color online) (a) Variation of the joint photon-phonon
number distribution for Stokes-phonon mode is shown with photon/phonon
numbers in Stokes and phonon (as $n_{S}=n_{V}$) modes and frequency
detuning parameter $\Delta\omega_{1}$ at rescaled time $gt=0.1$.
(b) Variation of the joint photon-phonon number distribution
for Stokes-phonon mode is shown with the frequency detuning parameter
$\Delta\omega_{1}$ and rescaled time for $n_{S}=n_{V}=0.1$. (c)
The dependence of the joint photon-phonon number distribution of the
pump-phonon mode on two frequency detuning parameters considering
$n_{V}=0.12$, $n_{L}=0.06$, and $gt=0.1$. (d) Joint photon-phonon
number distribution of the pump-phonon mode is shown as a function
of  the pump photon number and detuning $\Delta\omega_{1}=\pm\Delta\omega_{2}=\Delta\omega$ considering
$n_{V}=2$, and $gt=0.1$. Here, we have also assumed $I_{L}=10,\,I_{A}=1,$ and
$\chi=g$.}
\end{figure}

\begin{figure}[ht]
\begin{centering}
\includegraphics[scale=0.5]{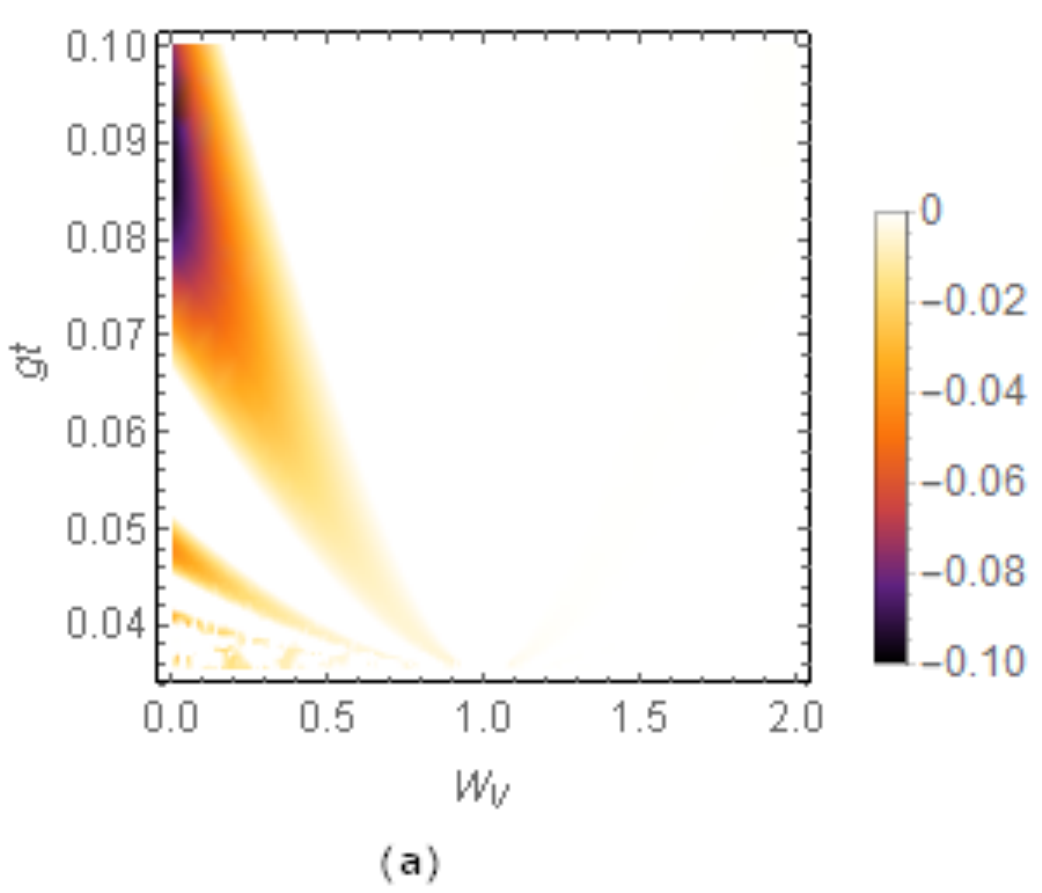} \includegraphics[scale=0.5]{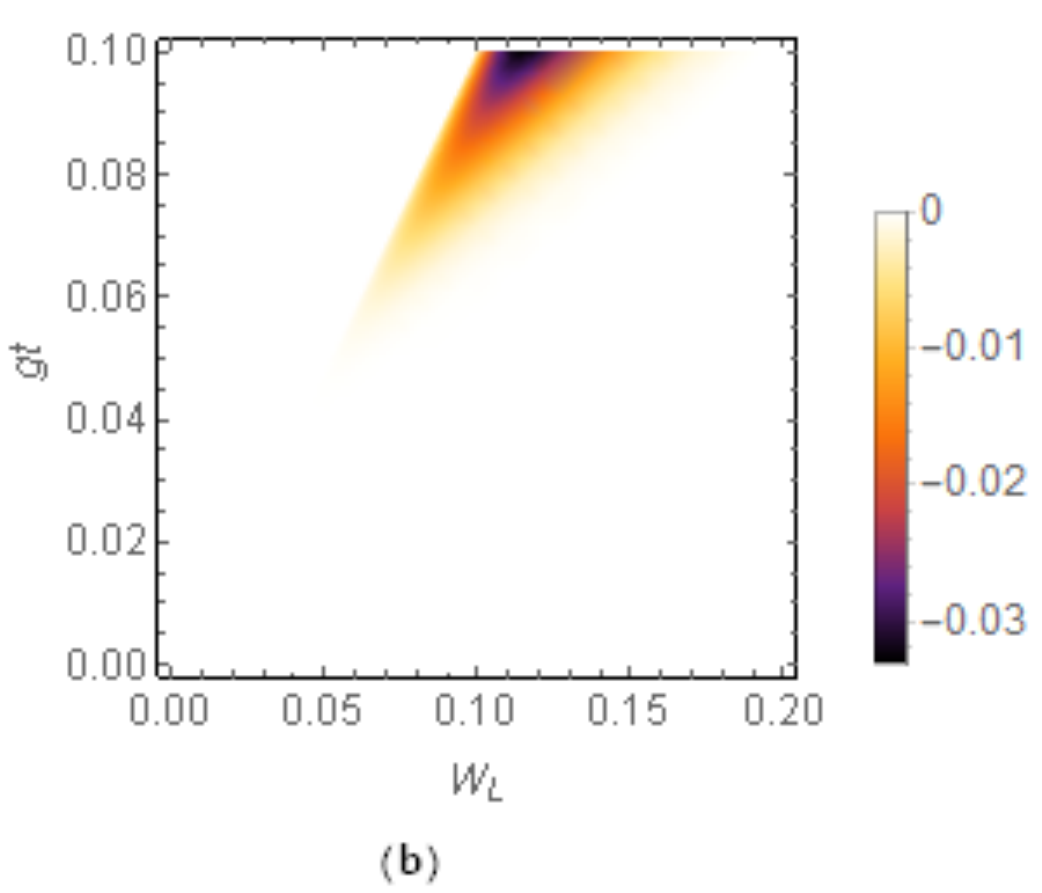}
\par\end{centering}
\caption{\label{fig:Int-inten}(Color online) Time evolution of the nonclassical
features is illustrated through the negative parts of the $s$ ordered
integrated intensity distribution with $s=0.8$ and $s=1$ for (a) Stokes-phonon
and (b) pump-phonon modes, respectively. To obtain this
plot, we have assumed $I_{L}=10,\,I_{A}=1$ with $\chi=g$, and frequency
detuning $\Delta\omega_{1}=\pm\Delta\omega_{2}=g$ with
$W_{S}=1$ and $W_{V}=0.5$. }
\end{figure}

We will further discuss the joint photon-phonon number distribution
and show its dependence on the various parameters. Note that the distributions of Stokes-phonon mode depend only on the detuning parameter in Stokes generation, while pump-phonon distributions are functions of both detuning parameters. Specifically, as already mentioned for the
resonance conditions, our results match with those for short-time
case \cite{pathak2013nonclassicality}. Here, we observe that in case of Stokes-phonon mode, joint
photon-phonon number distribution decreases with frequency detuning
in Stokes generation and for the higher values of the number of photons/phonons
(cf. Fig. \ref{fig:joint-det} (a)). This is consistent with the previous
results where the same behavior with photon/phonon number was observed
in frequency matched condition \cite{pathak2013nonclassicality}.
For such small photon/phonon numbers, we have observed from the time
evolution of the joint photon-phonon number distribution that the
distribution has a maximum value at zero detuning (see Fig. \ref{fig:joint-det}
(b)). Similarly, we have shown in Fig. \ref{fig:joint-det} (c) that
the joint photon-phonon number distribution for pump-phonon mode is 
affected dominantly due to frequency mismatch in anti-Stokes generation 
than that in Stokes process. For $\Delta\omega_{1}$,
the maximum value of distribution is obtained for non-zero detuning,
while distribution is maximum for $\Delta\omega_{2}=0$. { The effect
of frequency detuning in Stokes/anti-Stokes generation is further discussed with
the dependence of the joint distribution on the number of pump photons
in Fig. \ref{fig:joint-det} (d). A particularly interesting choice of parameters $\Delta\omega_{1}=\Delta\omega_{2}$ corresponds to conservation of radiation energy as $2\omega_{L}=\omega_{S}+\omega_{A},$ and $\Delta\omega_{1}=-\Delta\omega_{2}$ leads to vibration excitation $2\omega_{V}=\omega_{A}-\omega_{V}.$ In the present case, we observed that the results are same in these two special conditions.}

\begin{figure}[ht]
\begin{centering}
\includegraphics[scale=0.5]{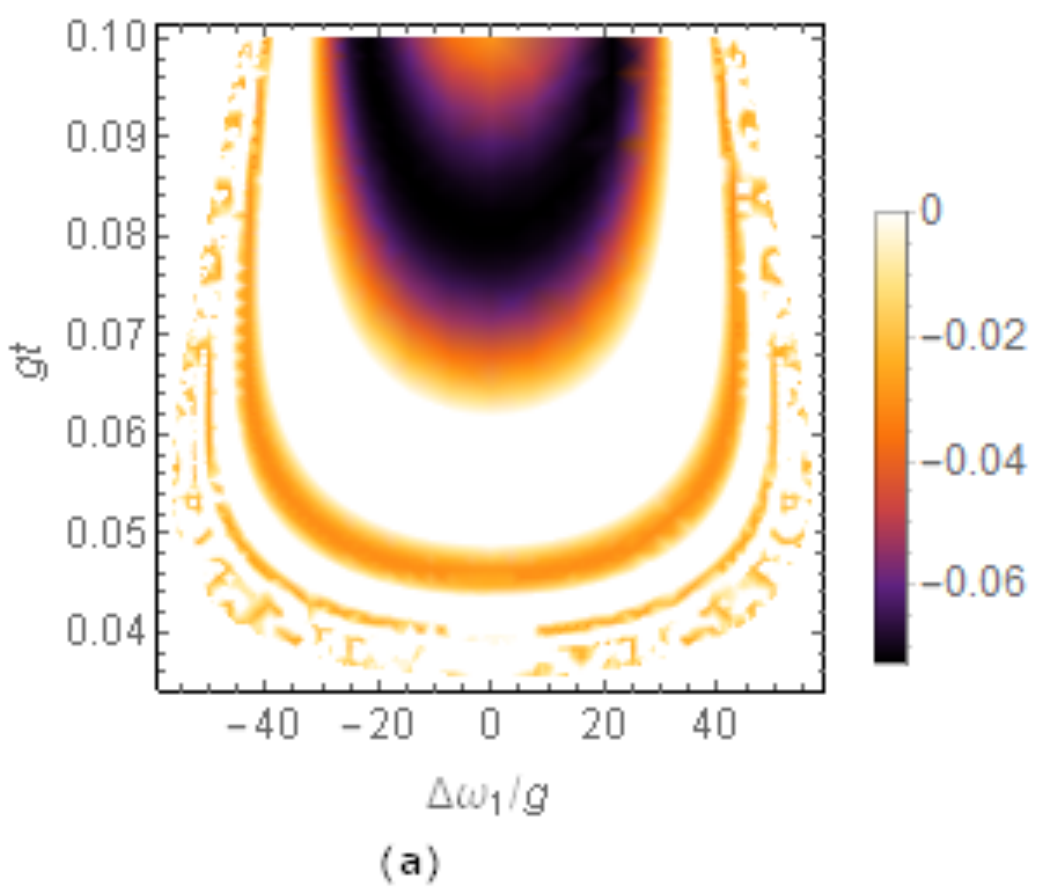} \includegraphics[scale=0.5]{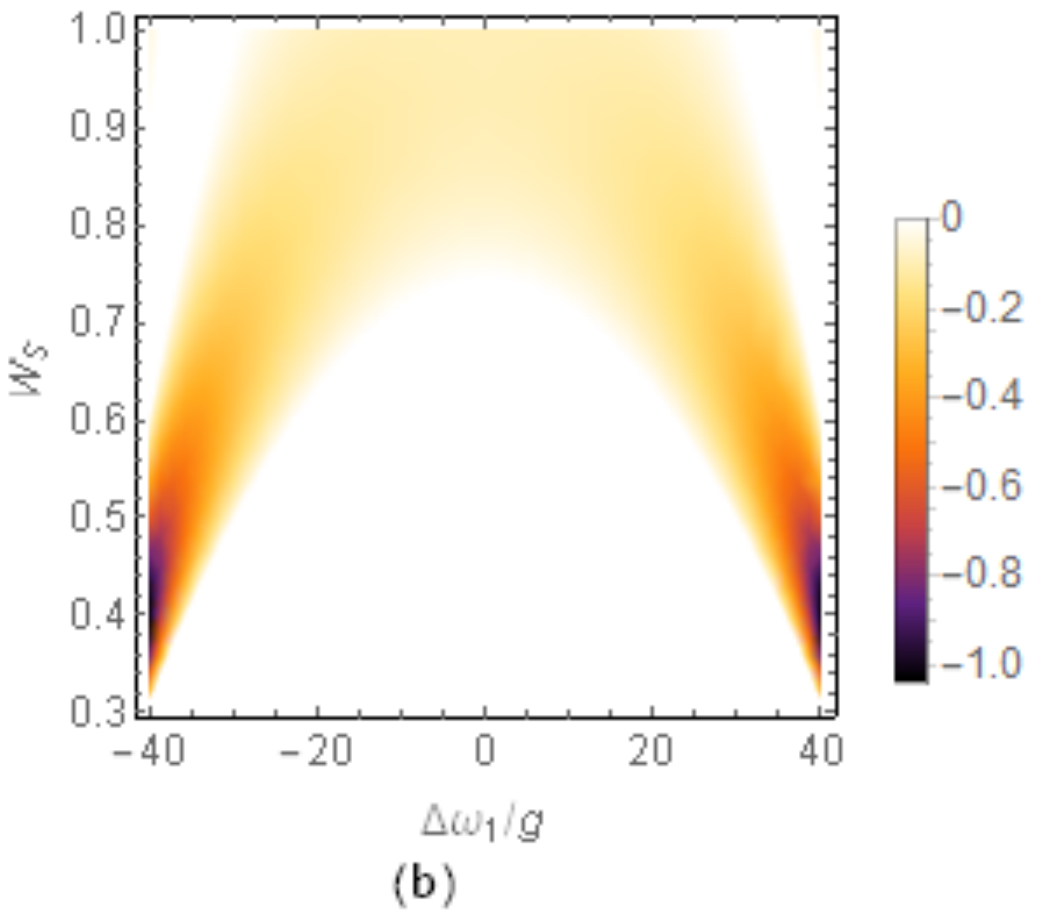} \includegraphics[scale=0.5]{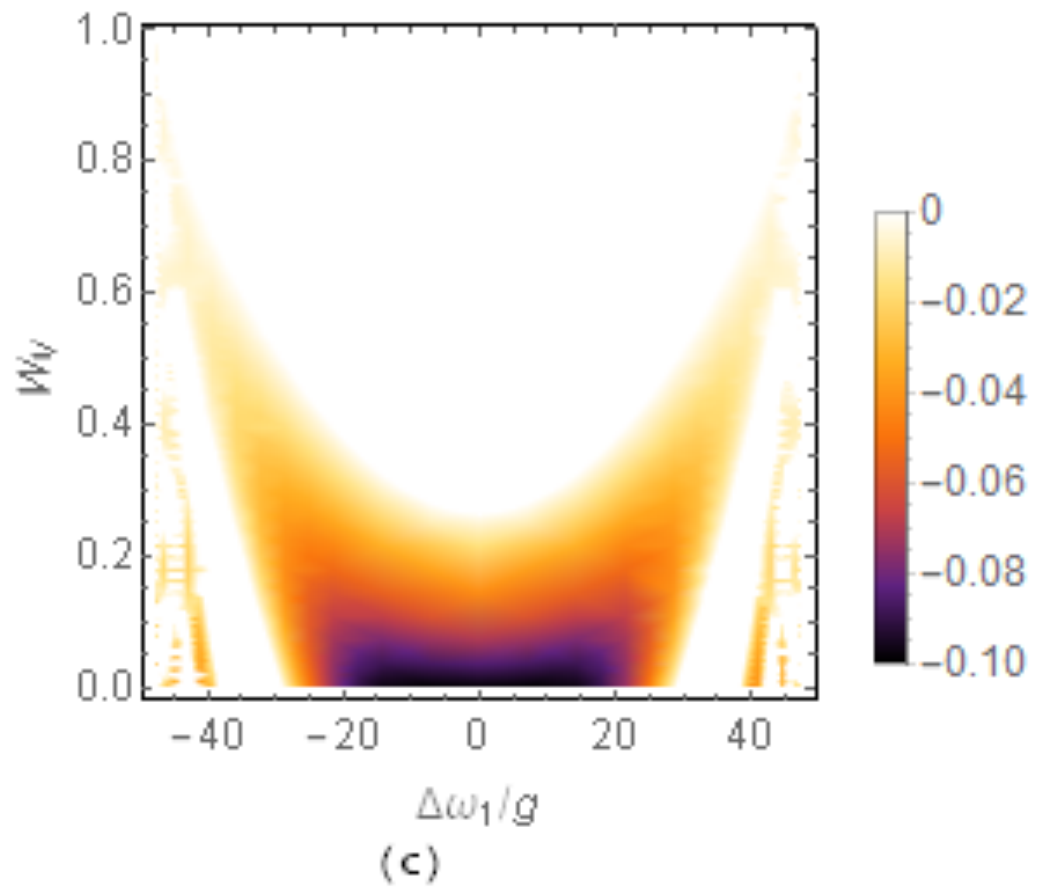} 
\par\end{centering}
\caption{\label{fig:Int-inten-det}(Color online) Nonclassical features are
illustrated through the negative parts of the $s$ ordered Stokes-phonon integrated
intensity distribution with $s=0.8$. Integrated intensity
distribution is shown to vary with frequency detuning $\Delta\omega_{1}$
and different parameters, as (a) rescaled time, (b) $W_{S}$, and
(c) $W_{V}$, considering $gt=0.09,$ $W_{S}=1$ and $W_{V}=0.1$
(in (a)) and 0.01 (in (b)). We have also assumed $I_{L}=10$ and $\chi=g$.}
\end{figure}

\begin{figure}
\begin{centering}
\includegraphics[scale=0.5]{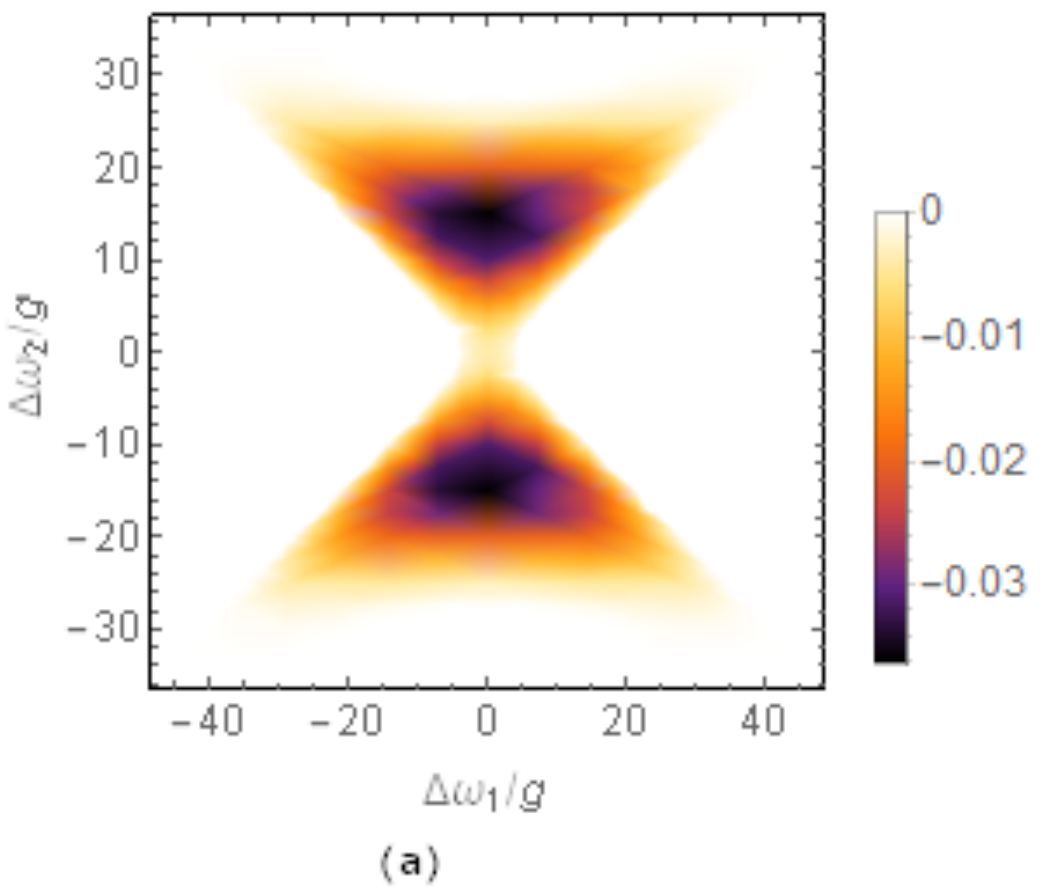} \includegraphics[scale=0.5]{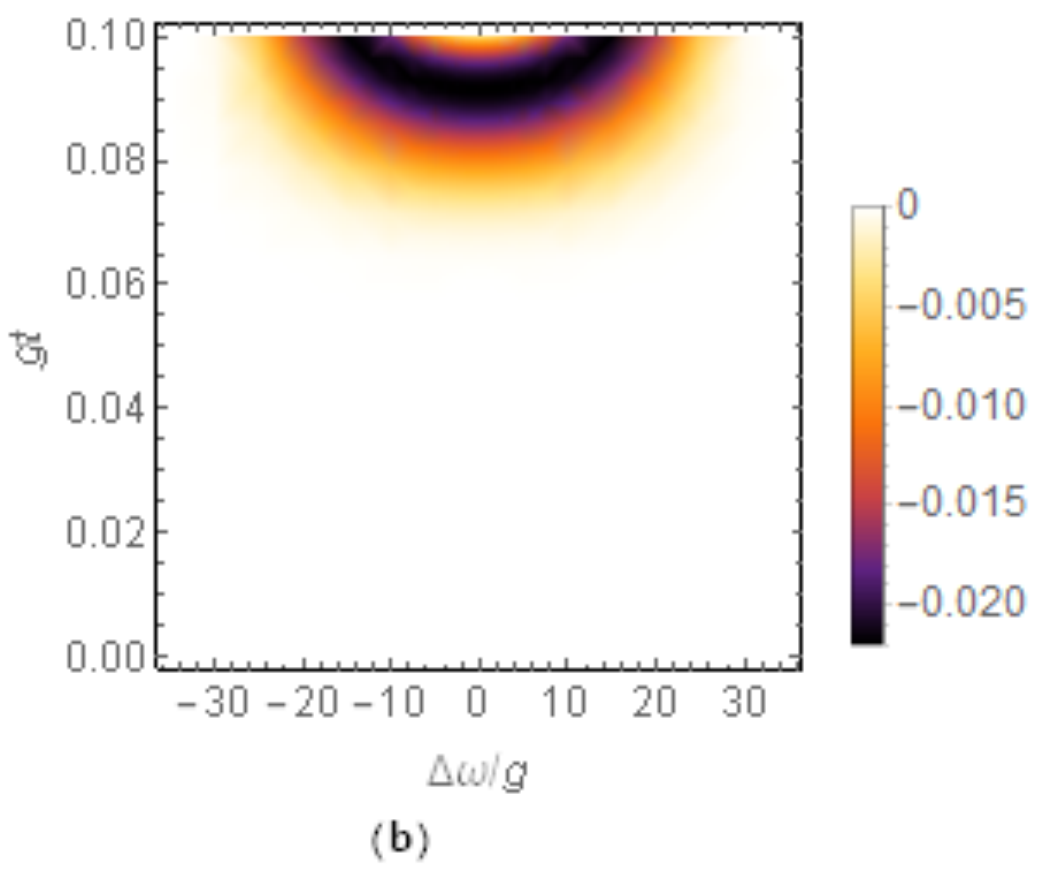} \includegraphics[scale=0.5]{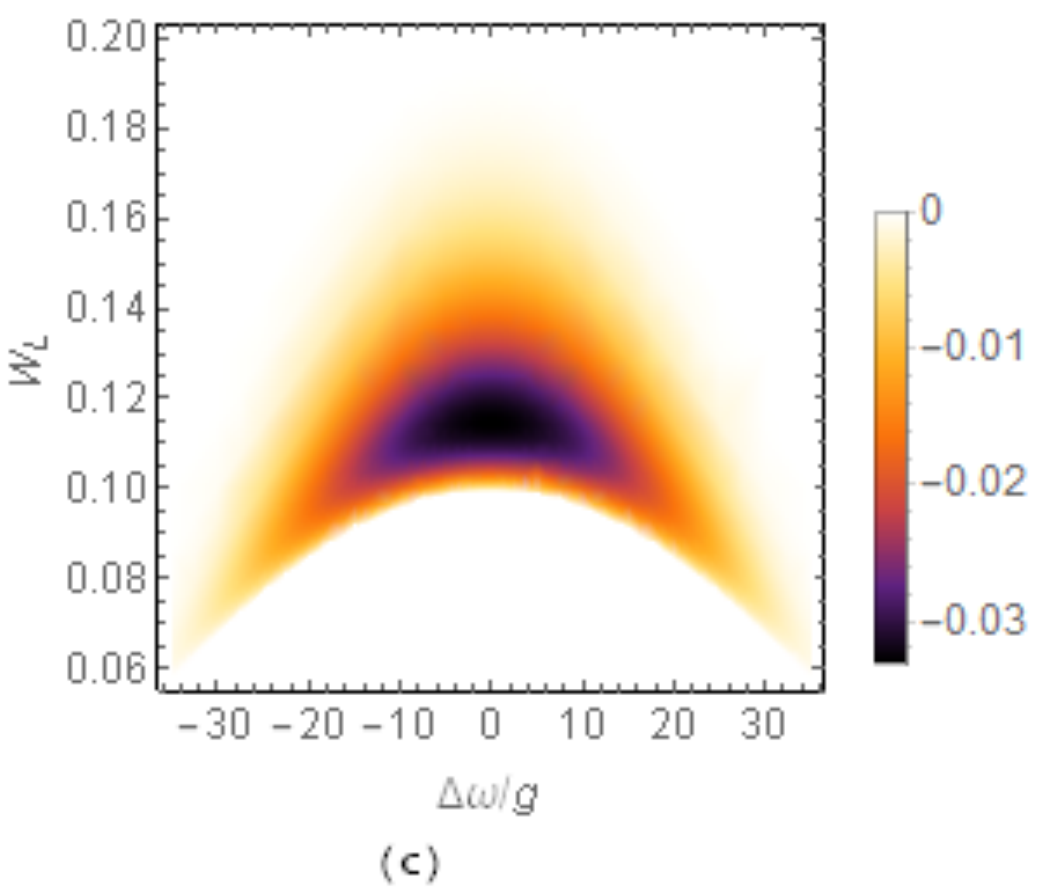} \includegraphics[scale=0.5]{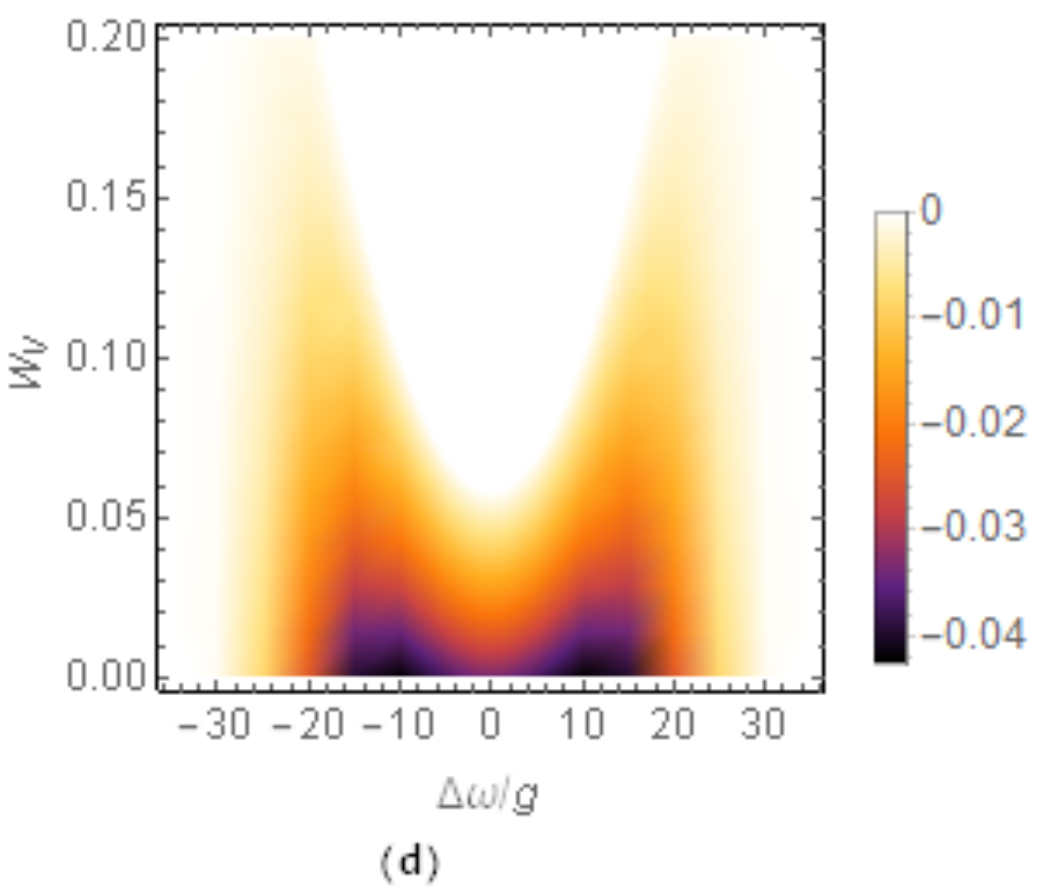}
\par\end{centering}
\caption{\label{fig:Int-inten-det2}(Color online) Nonclassical features are
illustrated through the negative parts of the Glauber-Sudarshan integrated
intensity distribution for pump-phonon mode as a function of intensity
and frequency detuning.
 (a) The dependence of the integrated intensity
distribution for pump-phonon mode on both the detuning parameters
is shown. (b)  Time evolution of the
nonclassicality present in this case as a function of frequency detuning
parameter  $\Delta\omega=\Delta\omega_{1}=\pm\Delta\omega_{2}$ considering $W_{L}=0.1$ and $W_{V}=0.05$.
The negative values of the distribution
as a function of frequency detuning
parameter  $\Delta\omega=\Delta\omega_{1}=\pm\Delta\omega_{2}$ and (c) $W_{L}$ with $W_{V}=0.05$,
and (d) $W_{V}$ with $W_{L}=1$. We have also assumed $I_{L}=10,\,I_{A}=1$ with $\chi=g$.}
\end{figure}

Further, nonclassicality reflected through the negative values of
the integrated intensity distributions is observed in both Stokes-phonon
and pump-phonon modes. Specifically, time evolution of integrated
intensity distribution in both these cases shows that nonclassicality
increases with interaction time (cf. Fig. \ref{fig:Int-inten}). Stokes-phonon
intensity distribution becomes negative for smaller values of phonon
intensity (cf. Fig. \ref{fig:Int-inten} (a)), while pump-phonon intensity
distribution shows nonclassicality only in specific case (cf. Fig.
\ref{fig:Int-inten} (b)). This nonclassical feature is further discussed
to analyze the effect of frequency detuning on the observed behavior
in Figs. \ref{fig:Int-inten-det} and \ref{fig:Int-inten-det2}. Specifically,
we have observed that it takes some time for the Stokes-phonon intensity
distribution to show nonclassical feature, which shows the similar
variation for a wide range of frequency detuning (cf. Fig. \ref{fig:Int-inten-det}
(a)). In general, the smaller values of both intensity of phonon mode
and frequency detuning are preferred to observe nonclassicality (cf.
Fig. \ref{fig:Int-inten-det} (c)). However, larger values of $\Delta\omega_{1}$
show nonclassicality with lower values of Stokes intensity (cf. Fig.
\ref{fig:Int-inten-det} (b)).

A similar study for pump-phonon
intensity distribution reveals that small (large) values of frequency
detuning in Stokes (anti-Stokes) process are preferred for the generation
of this nonclassicality in pump-phonon intensity distribution (shown
in Fig. \ref{fig:Int-inten-det2} (a)). Independently, time evolution
of the negative region of the integrated intensity distribution for
pump-phonon mode illustrates for photon and phonon frequency matching conditions (i.e., $\Delta\omega=\Delta\omega_{1}=\pm\Delta\omega_{2}$) that at larger values of the rescaled time, nonclassicality is generated for intermediate values of frequency detuning in Fig. \ref{fig:Int-inten-det2}
(b). 
Further analysis of the dependence of integrated intensity distribution
on the intensity of pump and phonon modes (in Fig. \ref{fig:Int-inten-det2}
(c) and (d)) shows that for lower intensity of pump (phonon) mode nonclassical
features are dominant for smaller (larger) values of $\Delta\omega$. The nonclassicality is observed for very small values of intensity of phonon than
that of the pump mode. 

\begin{figure}[ht]
\begin{centering}
\includegraphics[scale=0.4]{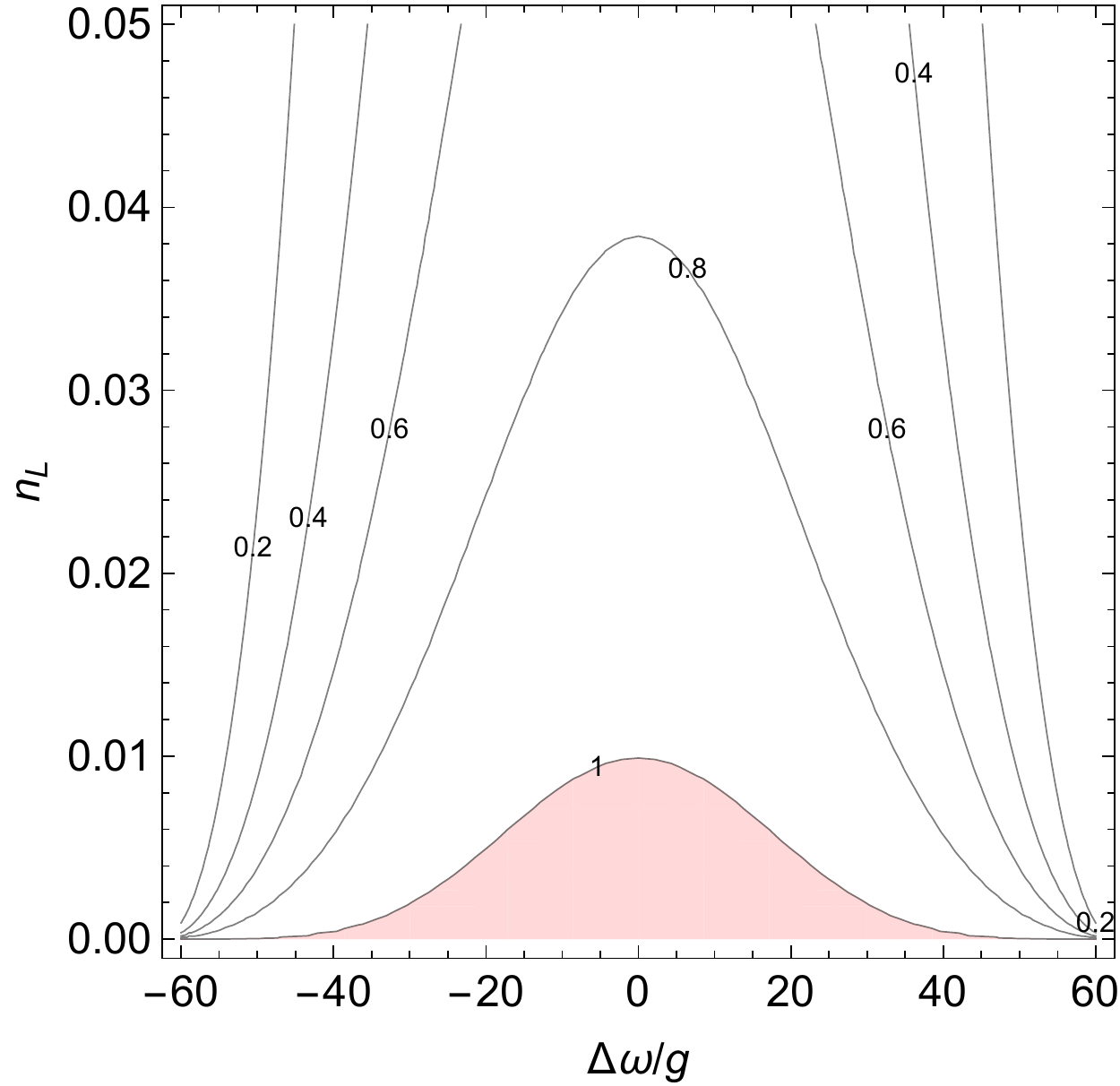}
\par\end{centering}
\caption{\label{fig:FVC}(Color online) Conditional Fano factor $F_{V,c}$
reflecting nonclassical behavior for values less than 1 is shown as
a function of $n_{L}$ and $\Delta\omega=\Delta\omega_{1}=\pm\Delta\omega_{2}$. The parameter shows
nonclassical feature except for the shaded region. We have chosen
$gt=0.1$, $I_{L}=10,\,I_{A}=1,$ and $\chi=g$.}
\end{figure}

\section{Nonclassicality using joint number distribution: Difference and conditional number distributions \label{sec:cond}}

The joint photon number distribution obtained in the previous section
also allows us to study nonclassical features. For instance, the conditional
Fano factor defined as 
\begin{equation}
F_{i,c}=\frac{\left\langle \left(\Delta n_{i}\right)^{2}\right\rangle _{c}}{\left\langle n_{i}\right\rangle },\label{eq:Con-Fano}
\end{equation}
which shows nonclassicality for $F_{i,c}<1$. As Stokes and phonon
modes are already shown to be generated in pair, we discuss here pump-phonon
mode only. In this case, conditional Fano factors for the pump and phonon
modes are 
\begin{equation}
F_{L,c}=1-\frac{B_{L}}{B_{V}}\label{eq:Con-Fano-pump}
\end{equation}
and 
\begin{equation}
F_{V,c}=\frac{\left(n_{L}+1\right)\left(\frac{1+B_{V}}{1+B_{L}}\right)^{2}-1}{\left(n_{L}+1\right)\left(\frac{1+B_{V}}{1+B_{L}}\right)-1}-1,\label{eq:Con-Fano-phon}
\end{equation}
respectively. We can clearly observe that pump mode shows signatures
of this nonclassical feature for any value of frequency mismatches as the quantity (\ref{eq:Con-Fano-pump}) is always less than
unity (it always holds here that $B_L<B_V$).
In case of the phonon mode, we have shown the conditional Fano factor
as a function of frequency detuning $\Delta\omega$ in Fig. \ref{fig:FVC} for equal or opposite photon and phonon detuning and have shown
that nonclassicality can be observed in general, but for small values
of frequency detuning and photon numbers in the pump mode. Specifically,
in this case as well, we can observe that with the increase in frequency
detuning the nonclassicality present in the phonon mode becomes prominent. 

\begin{figure}
\begin{centering}
\includegraphics[scale=0.5]{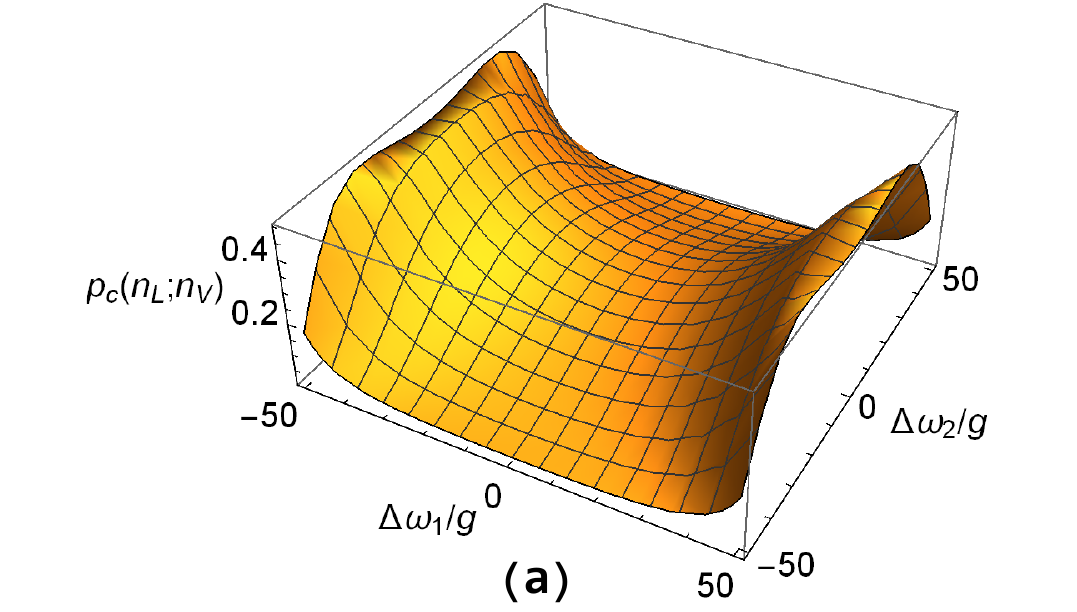} \includegraphics[scale=0.5]{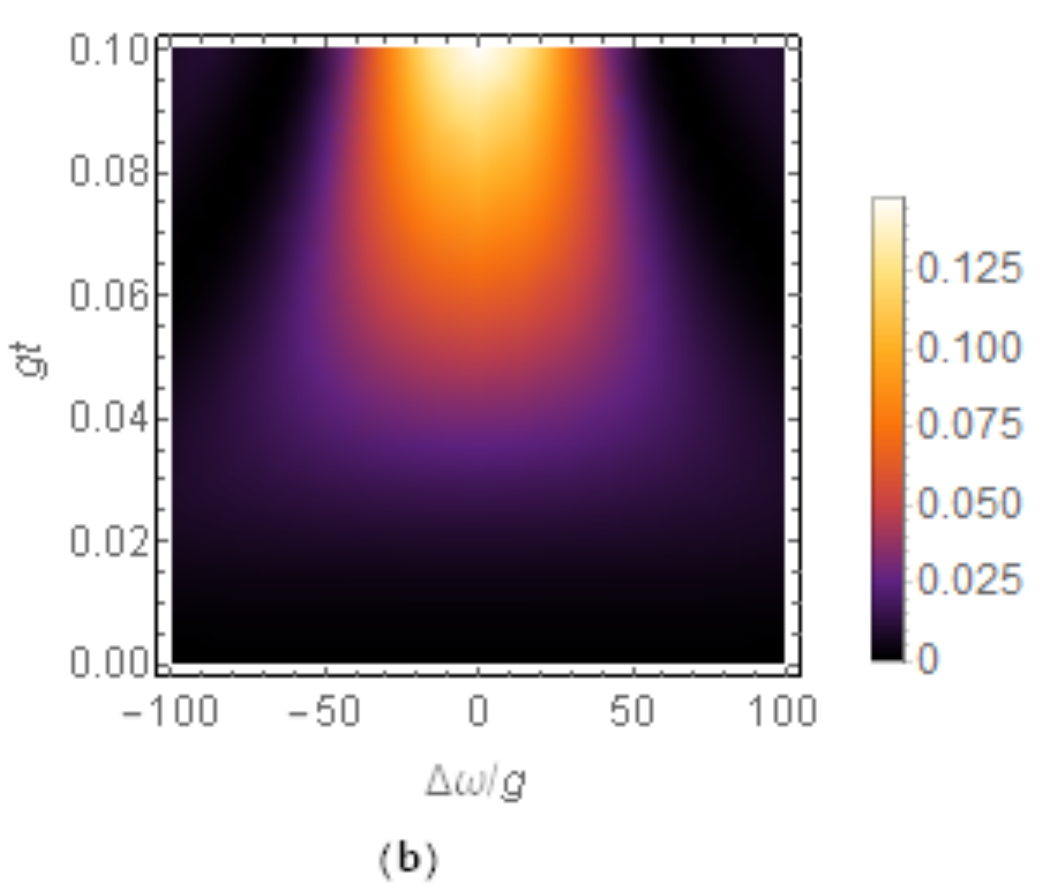}
\par\end{centering}
\caption{\label{fig:Cond-num}(Color online) Conditional photon number distribution
(a) $p_{c}\left(n_{L};n_{V}\right)$ and (b) $p_{c}\left(n_{V};n_{L}\right)$
are shown here as functions of frequency detuning. For $p_{c}\left(n_{L};n_{V}\right)$
the effect of both $\Delta\omega_{i}$s is discussed, while the time evolution
of $p_{c}\left(n_{V};n_{L}\right)$ as a function of $\Delta\omega=\Delta\omega_{1}=\pm\Delta\omega_{2}$
is shown. We have chosen $gt=0.1,$ $n_{V}=2,$
 $n_{L}=1$, and $I_{L}=10,\,I_{A}=1$ with $\chi=g$.}
\end{figure}

We have also obtained corresponding conditional number distributions as (using
Eq. 25 of \cite{perina2011joint})
\begin{equation}
\begin{array}{lcl}
p_{c}\left(n_{L};n_{V}\right) & = & \frac{n_{V}!}{n_{L}!\left(n_{V}!-n_{L}!\right)}\left(1-\frac{B_{L}}{B_{V}}\right)^{n_{V}}\left(\frac{B_{L}}{B_{V}-B_{L}}\right)^{n_{L}},\\
p_{c}\left(n_{V};n_{L}\right) & = & \frac{n_{V}!}{n_{L}!\left(n_{V}!-n_{L}!\right)}\left(\frac{1+B_{L}}{1+B_{V}}\right)\left(\frac{B_{V}-B_{L}}{1+B_{V}}\right)^{n_{V}}\\
& \times &\left(\frac{1+B_{L}}{B_{V}-B_{L}}\right)^{n_{L}}.
\end{array}\label{eq:Con-num}
\end{equation}
This provides declination from ideal diagonal distribution (that we
obtained for Stokes-phonon joint number distribution in Eq. (\ref{eq:pnsv})),
which is shown as a function of frequency detuning parameters in Fig.
\ref{fig:Cond-num}. We can observe in Fig.
\ref{fig:Cond-num} (a) that smaller values of $\Delta\omega_{2}$
and larger values of $\Delta\omega_{1}$ give higher values of conditional
photon number in pump mode (similar to Fig. \ref{fig:joint-det} (c)). Similarly, time evolution of conditional
phonon number distribution in Fig. \ref{fig:Cond-num} (b) shows that lower values of frequency detuning
$\Delta\omega$ at photon and phonon matching provide higher values to this parameter  (as in Fig. \ref{fig:joint-det} (d)).

\begin{figure}[h]
\begin{centering}
\includegraphics[scale=0.5]{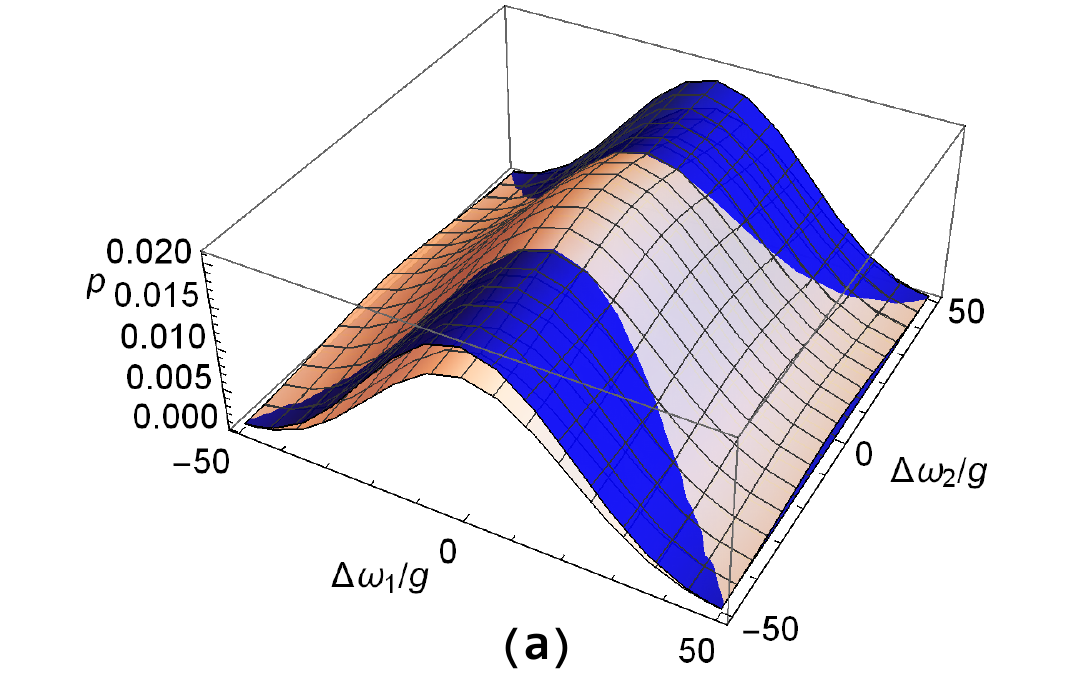} \includegraphics[scale=0.5]{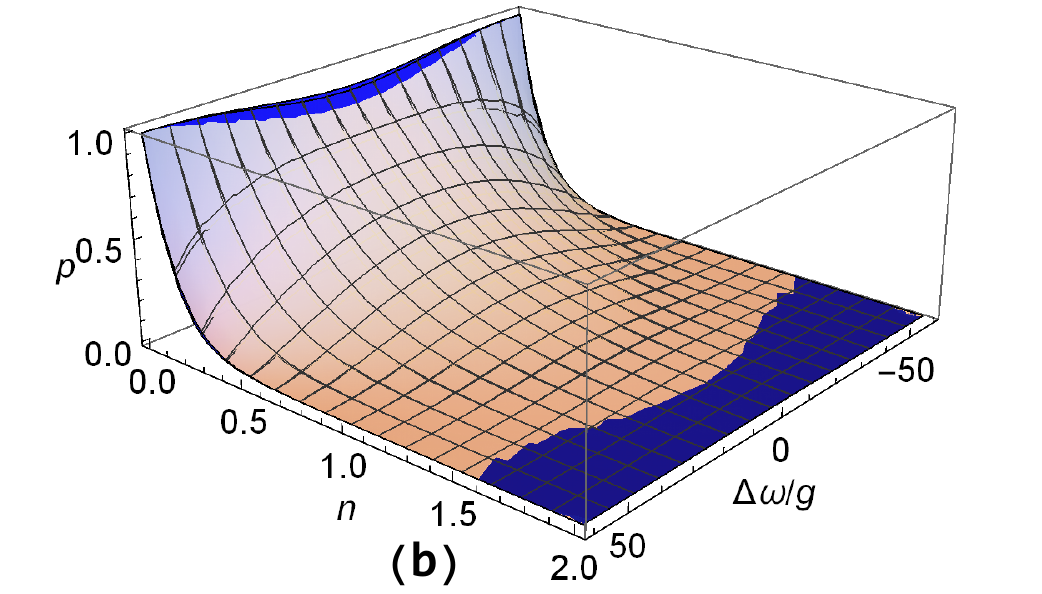}
\par\end{centering}
\caption{\label{fig:poiss}(Color online) Difference number distribution $p_{-}\left(n\right)$
and Poissonian distribution $p_{\textrm{Poiss}}\left(n\right)$ are
shown as functions of frequency detuning $\Delta\omega_{i}$ as dark
blue and light pink surfaces, respectively. We have chosen $gt=0.1$, $I_{L}=10,\,I_{A}=1,$ and $\chi=g$. In 
(a), we have considered $n=1.6$, while in (b), $\Delta\omega_{1}=\pm\Delta\omega_{2}=\Delta\omega$. }
\end{figure}

To characterize the nonclassical features and the quality of photon-phonon
pair generation in pump-phonon case, we have computed the difference number distribution
(using Eq. 26 of \cite{perina2011joint}) as 
\begin{equation}
p_{-}\left(n\right)=\frac{\left(B_{V}-B_{L}\right)^{n}}{\left(1+B_{V}-B_{L}\right)^{n+1}},\label{eq:diff}
\end{equation}
while Poissonian distribution for the same two combined modes will be 
\begin{equation}
p_{\textrm{Poiss}}\left(n\right)=\frac{\left(B_{V}+B_{L}\right)^{n}}{n!}\exp\left(-B_{V}-B_{L}\right).\label{eq:poisson}
\end{equation}
{ A combined plot of these two quantities ((\ref{eq:diff}) and (\ref{eq:poisson}))
shows that with a proper choice of frequency detuning both sub-Poissonian
and super-Poissonian characters can be observed in Fig. \ref{fig:poiss} (a)
in the difference number distribution for the same set of the rest
of the parameters.  It is worth mentioning here that the choice of the parameters is such that the short-time solution \cite{pathak2013nonclassicality} shows super-Poissonian difference number distribution. We have also discussed the case when detuning in Stokes generation is same as that of anti-Stokes generation, i.e., in photon and phonon matching conditions, and observed that the difference number distribution remains sub-Possonian for smaller values of the number  of photons (cf. Fig. \ref{fig:poiss} (b)).  }

\section{Conclusions \label{sec:Conclusion}}

The characteristic function for the off-resonant Raman process is obtained and shown to be Gaussian for all values of frequency detuning. The obtained characteristic function is shown to be more general than corresponding characteristic function obtained from the short-time solution, which can be obtained in the limiting case by considering either resonance condition or short-time approximation. The present study supports the results reporting nonclassical behavior in the short-time approximation and establishes that the validity of the short-time solution may be in the larger domain of time than usual expectations. 

In general, resonance conditions are associated with the performance of a nonlinear process, and in the present case Stokes and anti-Stokes generations are expected to be high in that case. This fact is also reflected through the joint Stokes-phonon number distribution discussed here. In contrast, we have shown here that the single- and two-mode nonclassicality in photon and phonon modes can be induced in the off-resonant conditions. Specifically, the present results establish that phonon mode remains entangled with both pump and Stokes modes for arbitrary value of frequency detuning for both initial coherent and chaotic phonon conditions. Sub-shot noise is observed in all the cases when entanglement is observed; on top of that, anti-Stokes-phonon mode with chaotic phonons also shows this nonclassicality. The presence of single-mode and intermodal squeezing for non-zero detuning certainly establishes the advantage of off-resonant Raman process in the generation of nonclassical states. Squeezing in the pump mode is favored by the higher values of frequency detuning in the Stokes generation for chaotic phonon. The pump mode is also obtained to be antibunched for both coherent and chaotic photons. On top of that, antibunching of phonon mode due to non-zero detuning may have applications in cavity optomechanics. The presence of two-mode correlations revealed through sum- and difference-variances is also complemented by the non-zero frequency detuning. We have summarized the nonclassical features that can be induced due to non-zero detuning in the off-resonant Raman process with coherent phonons in Table \ref{tab:non}.

\begin{table*}[t]
\begin{tabular}{p{3.9cm} p{5cm} p{5cm}}
Nonclassical feature & Present in Raman process at resonance & Observed additionally in off-resonant Raman process \tabularnewline
\hline 
Squeezing and antibunching & pump & phonon\tabularnewline
Intermodal squeezing & pump-phonon, pump-anti-Stokes, Stokes-phonon & pump-Stokes, Stokes-anti-Stokes, pump-anti-Stokes\tabularnewline
Sum and difference variance & pump-phonon, pump-anti-Stokes, Stokes-phonon, pump-Stokes & Stokes-anti-Stokes, pump-anti-Stokes\tabularnewline
Entanglement and sub-shot noise & pump-phonon, Stokes phonon & --\tabularnewline
\end{tabular}
{\caption{\label{tab:non} Summary of the nonclassical features observed in the off-resonant Raman process in addition of the Raman process at resonance considering all modes initially coherent.}}
\end{table*}

Subsequently, the joint photon number distribution for the pump-phonon and Stokes-phonon modes are obtained, which can be used to verify the quality of Stokes-phonon and pump-phonon pair generations with the help of conditional and difference photon/phonon number distributions. The joint pump-phonon number distribution is higher at the Stokes resonance condition which is an outcome of higher rate of Stokes generation. In contrast, joint pump-phonon number distribution prefers non-zero frequency detuning in Stokes process and frequency matching in anti-Stokes process, which can be attributed to the fact that a higher number of anti-Stokes photons annihilate to regenerate the pump-phonon pairs at anti-Stokes resonance. The nonclassicality reflected through conditional Fano factor, sub-Poissonian behavior of the difference number distribution is supported by the nonclassicality illustrated with the help of integrated intensity distributions. In both these cases, advantage of off-resonant Raman process is clearly visible.  Particularly, a variation of the threshold parameter establishes that the quantum features shown by integrated intensities can be enhanced by controlling the frequency detuning. The effect of phase matching conditions in the generation of nonclassicality in the off-resonant Raman process may also lead to interesting results. Further, off-resonant hyper-Raman process is also expected to give non-Gaussian characteristic function and several interesting nonclassical features.

We hope that analogous to the advantages in the field of quantum information processing, which are exploiting the facts originally thought to be the limitation of quantum theory, the present results on the off-resonant Raman process find applications through the generation of nonclassical states in these conditions. With recent improvements in experimental facilities to control detuning at the single photon level, we expect the present results to be useful there.

\textbf{Acknowledgement:} Authors thank the project LO1305 of the Ministry
of Education, Youth and Sports of the Czech Republic for support.

\bibliographystyle{apsrev4-1}
\bibliography{hyRa}

\begin{thebibliography}{53}%
\makeatletter
\providecommand \@ifxundefined [1]{%
 \@ifx{#1\undefined}
}%
\providecommand \@ifnum [1]{%
 \ifnum #1\expandafter \@firstoftwo
 \else \expandafter \@secondoftwo
 \fi
}%
\providecommand \@ifx [1]{%
 \ifx #1\expandafter \@firstoftwo
 \else \expandafter \@secondoftwo
 \fi
}%
\providecommand \natexlab [1]{#1}%
\providecommand \enquote  [1]{``#1''}%
\providecommand \bibnamefont  [1]{#1}%
\providecommand \bibfnamefont [1]{#1}%
\providecommand \citenamefont [1]{#1}%
\providecommand \href@noop [0]{\@secondoftwo}%
\providecommand \href [0]{\begingroup \@sanitize@url \@href}%
\providecommand \@href[1]{\@@startlink{#1}\@@href}%
\providecommand \@@href[1]{\endgroup#1\@@endlink}%
\providecommand \@sanitize@url [0]{\catcode `\\12\catcode `\$12\catcode
  `\&12\catcode `\#12\catcode `\^12\catcode `\_12\catcode `\%12\relax}%
\providecommand \@@startlink[1]{}%
\providecommand \@@endlink[0]{}%
\providecommand \url  [0]{\begingroup\@sanitize@url \@url }%
\providecommand \@url [1]{\endgroup\@href {#1}{\urlprefix }}%
\providecommand \urlprefix  [0]{URL }%
\providecommand \Eprint [0]{\href }%
\providecommand \doibase [0]{http://dx.doi.org/}%
\providecommand \selectlanguage [0]{\@gobble}%
\providecommand \bibinfo  [0]{\@secondoftwo}%
\providecommand \bibfield  [0]{\@secondoftwo}%
\providecommand \translation [1]{[#1]}%
\providecommand \BibitemOpen [0]{}%
\providecommand \bibitemStop [0]{}%
\providecommand \bibitemNoStop [0]{.\EOS\space}%
\providecommand \EOS [0]{\spacefactor3000\relax}%
\providecommand \BibitemShut  [1]{\csname bibitem#1\endcsname}%
\let\auto@bib@innerbib\@empty
\bibitem [{\citenamefont {Glauber}(1963)}]{glauber1963coherent}%
  \BibitemOpen
  \bibfield  {author} {\bibinfo {author} {\bibfnamefont {R.~J.}\ \bibnamefont
  {Glauber}},\ }\href@noop {} {\bibfield  {journal} {\bibinfo  {journal} {Phys.
  Rev.}\ }\textbf {\bibinfo {volume} {131}},\ \bibinfo {pages} {2766} (\bibinfo
  {year} {1963})}\BibitemShut {NoStop}%
\bibitem [{\citenamefont {Sudarshan}(1963)}]{sudarshan1963equivalence}%
  \BibitemOpen
  \bibfield  {author} {\bibinfo {author} {\bibfnamefont {E.~C.~G.}\
  \bibnamefont {Sudarshan}},\ }\href@noop {} {\bibfield  {journal} {\bibinfo
  {journal} {Phys. Rev. Lett.}\ }\textbf {\bibinfo {volume} {10}},\ \bibinfo
  {pages} {277} (\bibinfo {year} {1963})}\BibitemShut {NoStop}%
\bibitem [{\citenamefont {Thapliyal}\ \emph
  {et~al.}(2014{\natexlab{a}})\citenamefont {Thapliyal}, \citenamefont
  {Pathak}, \citenamefont {Sen},\ and\ \citenamefont
  {Pe{\v{r}}ina}}]{thapliyal2014higher}%
  \BibitemOpen
  \bibfield  {author} {\bibinfo {author} {\bibfnamefont {K.}~\bibnamefont
  {Thapliyal}}, \bibinfo {author} {\bibfnamefont {A.}~\bibnamefont {Pathak}},
  \bibinfo {author} {\bibfnamefont {B.}~\bibnamefont {Sen}}, \ and\ \bibinfo
  {author} {\bibfnamefont {J.}~\bibnamefont {Pe{\v{r}}ina}},\ }\href@noop {}
  {\bibfield  {journal} {\bibinfo  {journal} {Phys. Rev. A}\ }\textbf {\bibinfo
  {volume} {90}},\ \bibinfo {pages} {013808} (\bibinfo {year}
  {2014}{\natexlab{a}})}\BibitemShut {NoStop}%
\bibitem [{\citenamefont {Thapliyal}\ \emph
  {et~al.}(2014{\natexlab{b}})\citenamefont {Thapliyal}, \citenamefont
  {Pathak}, \citenamefont {Sen},\ and\ \citenamefont
  {Pe{\v{r}}ina}}]{thapliyal2014nonclassical}%
  \BibitemOpen
  \bibfield  {author} {\bibinfo {author} {\bibfnamefont {K.}~\bibnamefont
  {Thapliyal}}, \bibinfo {author} {\bibfnamefont {A.}~\bibnamefont {Pathak}},
  \bibinfo {author} {\bibfnamefont {B.}~\bibnamefont {Sen}}, \ and\ \bibinfo
  {author} {\bibfnamefont {J.}~\bibnamefont {Pe{\v{r}}ina}},\ }\href@noop {}
  {\bibfield  {journal} {\bibinfo  {journal} {Phys. Lett. A}\ }\textbf
  {\bibinfo {volume} {378}},\ \bibinfo {pages} {3431} (\bibinfo {year}
  {2014}{\natexlab{b}})}\BibitemShut {NoStop}%
\bibitem [{\citenamefont {Zeng}\ \emph {et~al.}(1995)\citenamefont {Zeng},
  \citenamefont {Zhang},\ and\ \citenamefont {Lin}}]{zeng1995nonclassical}%
  \BibitemOpen
  \bibfield  {author} {\bibinfo {author} {\bibfnamefont {H.}~\bibnamefont
  {Zeng}}, \bibinfo {author} {\bibfnamefont {W.}~\bibnamefont {Zhang}}, \ and\
  \bibinfo {author} {\bibfnamefont {F.}~\bibnamefont {Lin}},\ }\href@noop {}
  {\bibfield  {journal} {\bibinfo  {journal} {Phys. Rev. A}\ }\textbf {\bibinfo
  {volume} {52}},\ \bibinfo {pages} {2155} (\bibinfo {year}
  {1995})}\BibitemShut {NoStop}%
\bibitem [{\citenamefont {Giri}\ \emph {et~al.}(2017)\citenamefont {Giri},
  \citenamefont {Thapliyal}, \citenamefont {Sen},\ and\ \citenamefont
  {Pathak}}]{giri2017nonclassicality}%
  \BibitemOpen
  \bibfield  {author} {\bibinfo {author} {\bibfnamefont {S.~K.}\ \bibnamefont
  {Giri}}, \bibinfo {author} {\bibfnamefont {K.}~\bibnamefont {Thapliyal}},
  \bibinfo {author} {\bibfnamefont {B.}~\bibnamefont {Sen}}, \ and\ \bibinfo
  {author} {\bibfnamefont {A.}~\bibnamefont {Pathak}},\ }\href@noop {}
  {\bibfield  {journal} {\bibinfo  {journal} {Physica A}\ }\textbf {\bibinfo
  {volume} {466}},\ \bibinfo {pages} {140} (\bibinfo {year}
  {2017})}\BibitemShut {NoStop}%
\bibitem [{\citenamefont {Naikoo}\ \emph
  {et~al.}(2018{\natexlab{a}})\citenamefont {Naikoo}, \citenamefont
  {Thapliyal}, \citenamefont {Pathak},\ and\ \citenamefont
  {Banerjee}}]{naikoo2018probing}%
  \BibitemOpen
  \bibfield  {author} {\bibinfo {author} {\bibfnamefont {J.}~\bibnamefont
  {Naikoo}}, \bibinfo {author} {\bibfnamefont {K.}~\bibnamefont {Thapliyal}},
  \bibinfo {author} {\bibfnamefont {A.}~\bibnamefont {Pathak}}, \ and\ \bibinfo
  {author} {\bibfnamefont {S.}~\bibnamefont {Banerjee}},\ }\href@noop {}
  {\bibfield  {journal} {\bibinfo  {journal} {Phys. Rev. A}\ }\textbf {\bibinfo
  {volume} {97}},\ \bibinfo {pages} {063840} (\bibinfo {year}
  {2018}{\natexlab{a}})}\BibitemShut {NoStop}%
\bibitem [{\citenamefont {Naikoo}\ \emph
  {et~al.}(2018{\natexlab{b}})\citenamefont {Naikoo}, \citenamefont
  {Thapliyal}, \citenamefont {Banerjee},\ and\ \citenamefont
  {Pathak}}]{naikoo2018quantum}%
  \BibitemOpen
  \bibfield  {author} {\bibinfo {author} {\bibfnamefont {J.}~\bibnamefont
  {Naikoo}}, \bibinfo {author} {\bibfnamefont {K.}~\bibnamefont {Thapliyal}},
  \bibinfo {author} {\bibfnamefont {S.}~\bibnamefont {Banerjee}}, \ and\
  \bibinfo {author} {\bibfnamefont {A.}~\bibnamefont {Pathak}},\ }\href@noop {}
  {\bibfield  {journal} {\bibinfo  {journal} {arXiv preprint arXiv:1811.05604}\
  } (\bibinfo {year} {2018}{\natexlab{b}})}\BibitemShut {NoStop}%
\bibitem [{\citenamefont {Brooks}\ \emph {et~al.}(2012)\citenamefont {Brooks},
  \citenamefont {Botter}, \citenamefont {Schreppler}, \citenamefont {Purdy},
  \citenamefont {Brahms},\ and\ \citenamefont {Stamper-Kurn}}]{brooks2012non}%
  \BibitemOpen
  \bibfield  {author} {\bibinfo {author} {\bibfnamefont {D.~W.}\ \bibnamefont
  {Brooks}}, \bibinfo {author} {\bibfnamefont {T.}~\bibnamefont {Botter}},
  \bibinfo {author} {\bibfnamefont {S.}~\bibnamefont {Schreppler}}, \bibinfo
  {author} {\bibfnamefont {T.~P.}\ \bibnamefont {Purdy}}, \bibinfo {author}
  {\bibfnamefont {N.}~\bibnamefont {Brahms}}, \ and\ \bibinfo {author}
  {\bibfnamefont {D.~M.}\ \bibnamefont {Stamper-Kurn}},\ }\href@noop {}
  {\bibfield  {journal} {\bibinfo  {journal} {Nature}\ }\textbf {\bibinfo
  {volume} {488}},\ \bibinfo {pages} {476} (\bibinfo {year}
  {2012})}\BibitemShut {NoStop}%
\bibitem [{\citenamefont {Alam}\ \emph {et~al.}(2017)\citenamefont {Alam},
  \citenamefont {Thapliyal}, \citenamefont {Pathak}, \citenamefont {Sen},
  \citenamefont {Verma},\ and\ \citenamefont {Mandal}}]{alam2017lower}%
  \BibitemOpen
  \bibfield  {author} {\bibinfo {author} {\bibfnamefont {N.}~\bibnamefont
  {Alam}}, \bibinfo {author} {\bibfnamefont {K.}~\bibnamefont {Thapliyal}},
  \bibinfo {author} {\bibfnamefont {A.}~\bibnamefont {Pathak}}, \bibinfo
  {author} {\bibfnamefont {B.}~\bibnamefont {Sen}}, \bibinfo {author}
  {\bibfnamefont {A.}~\bibnamefont {Verma}}, \ and\ \bibinfo {author}
  {\bibfnamefont {S.}~\bibnamefont {Mandal}},\ }\href@noop {} {\bibfield
  {journal} {\bibinfo  {journal} {arXiv preprint arXiv:1708.03967}\ } (\bibinfo
  {year} {2017})}\BibitemShut {NoStop}%
\bibitem [{\citenamefont {Wu}\ \emph {et~al.}(1986)\citenamefont {Wu},
  \citenamefont {Kimble}, \citenamefont {Hall},\ and\ \citenamefont
  {Wu}}]{wu1986generation}%
  \BibitemOpen
  \bibfield  {author} {\bibinfo {author} {\bibfnamefont {L.-A.}\ \bibnamefont
  {Wu}}, \bibinfo {author} {\bibfnamefont {H.}~\bibnamefont {Kimble}}, \bibinfo
  {author} {\bibfnamefont {J.}~\bibnamefont {Hall}}, \ and\ \bibinfo {author}
  {\bibfnamefont {H.}~\bibnamefont {Wu}},\ }\href@noop {} {\bibfield  {journal}
  {\bibinfo  {journal} {Phys. Rev. Lett.}\ }\textbf {\bibinfo {volume} {57}},\
  \bibinfo {pages} {2520} (\bibinfo {year} {1986})}\BibitemShut {NoStop}%
\bibitem [{\citenamefont {Pe{\v{r}}ina}\ and\ \citenamefont
  {K{\v{r}}epelka}(2011)}]{perina2011joint}%
  \BibitemOpen
  \bibfield  {author} {\bibinfo {author} {\bibfnamefont {J.}~\bibnamefont
  {Pe{\v{r}}ina}}\ and\ \bibinfo {author} {\bibfnamefont {J.}~\bibnamefont
  {K{\v{r}}epelka}},\ }\href@noop {} {\bibfield  {journal} {\bibinfo  {journal}
  {Opt. Comm.}\ }\textbf {\bibinfo {volume} {284}},\ \bibinfo {pages} {4941}
  (\bibinfo {year} {2011})}\BibitemShut {NoStop}%
\bibitem [{\citenamefont {Slusher}\ \emph {et~al.}(1985)\citenamefont
  {Slusher}, \citenamefont {Hollberg}, \citenamefont {Yurke}, \citenamefont
  {Mertz},\ and\ \citenamefont {Valley}}]{slusher1985observation}%
  \BibitemOpen
  \bibfield  {author} {\bibinfo {author} {\bibfnamefont {R.}~\bibnamefont
  {Slusher}}, \bibinfo {author} {\bibfnamefont {L.}~\bibnamefont {Hollberg}},
  \bibinfo {author} {\bibfnamefont {B.}~\bibnamefont {Yurke}}, \bibinfo
  {author} {\bibfnamefont {J.}~\bibnamefont {Mertz}}, \ and\ \bibinfo {author}
  {\bibfnamefont {J.}~\bibnamefont {Valley}},\ }\href@noop {} {\bibfield
  {journal} {\bibinfo  {journal} {Phys. Rev. Lett.}\ }\textbf {\bibinfo
  {volume} {55}},\ \bibinfo {pages} {2409} (\bibinfo {year}
  {1985})}\BibitemShut {NoStop}%
\bibitem [{\citenamefont {Pe{\v{r}}ina}\ and\ \citenamefont
  {K{\v{r}}epelka}(1992)}]{perina1992quantum}%
  \BibitemOpen
  \bibfield  {author} {\bibinfo {author} {\bibfnamefont {J.}~\bibnamefont
  {Pe{\v{r}}ina}}\ and\ \bibinfo {author} {\bibfnamefont {J.}~\bibnamefont
  {K{\v{r}}epelka}},\ }\href@noop {} {\bibfield  {journal} {\bibinfo  {journal}
  {J. Mod. Opt.}\ }\textbf {\bibinfo {volume} {39}},\ \bibinfo {pages} {1029}
  (\bibinfo {year} {1992})}\BibitemShut {NoStop}%
\bibitem [{\citenamefont {Thapliyal}\ \emph {et~al.}(2017)\citenamefont
  {Thapliyal}, \citenamefont {Pathak}, \citenamefont {Sen},\ and\ \citenamefont
  {Pe{\v{r}}ina}}]{thapliyal2017nonclassicality}%
  \BibitemOpen
  \bibfield  {author} {\bibinfo {author} {\bibfnamefont {K.}~\bibnamefont
  {Thapliyal}}, \bibinfo {author} {\bibfnamefont {A.}~\bibnamefont {Pathak}},
  \bibinfo {author} {\bibfnamefont {B.}~\bibnamefont {Sen}}, \ and\ \bibinfo
  {author} {\bibfnamefont {J.}~\bibnamefont {Pe{\v{r}}ina}},\ }\href@noop {}
  {\bibfield  {journal} {\bibinfo  {journal} {arXiv preprint arXiv:1710.04456}\
  } (\bibinfo {year} {2017})}\BibitemShut {NoStop}%
\bibitem [{\citenamefont {Pe{\v{r}}ina}(1991)}]{perina1991quantum}%
  \BibitemOpen
  \bibfield  {author} {\bibinfo {author} {\bibfnamefont {J.}~\bibnamefont
  {Pe{\v{r}}ina}},\ }\href@noop {} {\emph {\bibinfo {title} {Quantum Statistics
  of Linear and Nonlinear Optical Phenomena}}}\ (\bibinfo  {publisher} {Kluwer
  Academic, Dordrecht-Boston},\ \bibinfo {year} {1991})\BibitemShut {NoStop}%
\bibitem [{\citenamefont {Bennett}\ and\ \citenamefont
  {Brassard}(1984)}]{bennett1984quantum}%
  \BibitemOpen
  \bibfield  {author} {\bibinfo {author} {\bibfnamefont {C.~H.}\ \bibnamefont
  {Bennett}}\ and\ \bibinfo {author} {\bibfnamefont {G.}~\bibnamefont
  {Brassard}},\ }in\ \href@noop {} {\emph {\bibinfo {booktitle} {International
  Conference on Computer System and Signal Processing, IEEE, 1984}}}\ (\bibinfo
  {year} {1984})\ pp.\ \bibinfo {pages} {175--179}\BibitemShut {NoStop}%
\bibitem [{\citenamefont {Hillery}(2000)}]{hillery2000quantum}%
  \BibitemOpen
  \bibfield  {author} {\bibinfo {author} {\bibfnamefont {M.}~\bibnamefont
  {Hillery}},\ }\href@noop {} {\bibfield  {journal} {\bibinfo  {journal} {Phys.
  Rev. A}\ }\textbf {\bibinfo {volume} {61}},\ \bibinfo {pages} {022309}
  (\bibinfo {year} {2000})}\BibitemShut {NoStop}%
\bibitem [{\citenamefont {Ekert}(1991)}]{ekert1991quantum}%
  \BibitemOpen
  \bibfield  {author} {\bibinfo {author} {\bibfnamefont {A.~K.}\ \bibnamefont
  {Ekert}},\ }\href@noop {} {\bibfield  {journal} {\bibinfo  {journal} {Phys.
  Rev. Lett.}\ }\textbf {\bibinfo {volume} {67}},\ \bibinfo {pages} {661}
  (\bibinfo {year} {1991})}\BibitemShut {NoStop}%
\bibitem [{\citenamefont {Branciard}\ \emph {et~al.}(2012)\citenamefont
  {Branciard}, \citenamefont {Cavalcanti}, \citenamefont {Walborn},
  \citenamefont {Scarani},\ and\ \citenamefont {Wiseman}}]{branciard2012one}%
  \BibitemOpen
  \bibfield  {author} {\bibinfo {author} {\bibfnamefont {C.}~\bibnamefont
  {Branciard}}, \bibinfo {author} {\bibfnamefont {E.~G.}\ \bibnamefont
  {Cavalcanti}}, \bibinfo {author} {\bibfnamefont {S.~P.}\ \bibnamefont
  {Walborn}}, \bibinfo {author} {\bibfnamefont {V.}~\bibnamefont {Scarani}}, \
  and\ \bibinfo {author} {\bibfnamefont {H.~M.}\ \bibnamefont {Wiseman}},\
  }\href@noop {} {\bibfield  {journal} {\bibinfo  {journal} {Phys. Rev. A}\
  }\textbf {\bibinfo {volume} {85}},\ \bibinfo {pages} {010301} (\bibinfo
  {year} {2012})}\BibitemShut {NoStop}%
\bibitem [{\citenamefont {Acin}\ \emph {et~al.}(2006)\citenamefont {Acin},
  \citenamefont {Gisin},\ and\ \citenamefont {Masanes}}]{acin2006bell}%
  \BibitemOpen
  \bibfield  {author} {\bibinfo {author} {\bibfnamefont {A.}~\bibnamefont
  {Acin}}, \bibinfo {author} {\bibfnamefont {N.}~\bibnamefont {Gisin}}, \ and\
  \bibinfo {author} {\bibfnamefont {L.}~\bibnamefont {Masanes}},\ }\href@noop
  {} {\bibfield  {journal} {\bibinfo  {journal} {Phys. Rev. Lett.}\ }\textbf
  {\bibinfo {volume} {97}},\ \bibinfo {pages} {120405} (\bibinfo {year}
  {2006})}\BibitemShut {NoStop}%
\bibitem [{\citenamefont {Herrero-Collantes}\ and\ \citenamefont
  {Garcia-Escartin}(2017)}]{herrero2017quantum}%
  \BibitemOpen
  \bibfield  {author} {\bibinfo {author} {\bibfnamefont {M.}~\bibnamefont
  {Herrero-Collantes}}\ and\ \bibinfo {author} {\bibfnamefont {J.~C.}\
  \bibnamefont {Garcia-Escartin}},\ }\href@noop {} {\bibfield  {journal}
  {\bibinfo  {journal} {Rev. Mod. Phys.}\ }\textbf {\bibinfo {volume} {89}},\
  \bibinfo {pages} {015004} (\bibinfo {year} {2017})}\BibitemShut {NoStop}%
\bibitem [{\citenamefont {Giovannetti}\ \emph {et~al.}(2011)\citenamefont
  {Giovannetti}, \citenamefont {Lloyd},\ and\ \citenamefont
  {Maccone}}]{giovannetti2011advances}%
  \BibitemOpen
  \bibfield  {author} {\bibinfo {author} {\bibfnamefont {V.}~\bibnamefont
  {Giovannetti}}, \bibinfo {author} {\bibfnamefont {S.}~\bibnamefont {Lloyd}},
  \ and\ \bibinfo {author} {\bibfnamefont {L.}~\bibnamefont {Maccone}},\
  }\href@noop {} {\bibfield  {journal} {\bibinfo  {journal} {Nature photonics}\
  }\textbf {\bibinfo {volume} {5}},\ \bibinfo {pages} {222} (\bibinfo {year}
  {2011})}\BibitemShut {NoStop}%
\bibitem [{\citenamefont {Bennett}\ \emph {et~al.}(1993)\citenamefont
  {Bennett}, \citenamefont {Brassard}, \citenamefont {Cr{\'e}peau},
  \citenamefont {Jozsa}, \citenamefont {Peres},\ and\ \citenamefont
  {Wootters}}]{bennett1993teleporting}%
  \BibitemOpen
  \bibfield  {author} {\bibinfo {author} {\bibfnamefont {C.~H.}\ \bibnamefont
  {Bennett}}, \bibinfo {author} {\bibfnamefont {G.}~\bibnamefont {Brassard}},
  \bibinfo {author} {\bibfnamefont {C.}~\bibnamefont {Cr{\'e}peau}}, \bibinfo
  {author} {\bibfnamefont {R.}~\bibnamefont {Jozsa}}, \bibinfo {author}
  {\bibfnamefont {A.}~\bibnamefont {Peres}}, \ and\ \bibinfo {author}
  {\bibfnamefont {W.~K.}\ \bibnamefont {Wootters}},\ }\href@noop {} {\bibfield
  {journal} {\bibinfo  {journal} {Phys. Rev. Lett.}\ }\textbf {\bibinfo
  {volume} {70}},\ \bibinfo {pages} {1895} (\bibinfo {year}
  {1993})}\BibitemShut {NoStop}%
\bibitem [{\citenamefont {Bennett}\ and\ \citenamefont
  {Wiesner}(1992)}]{bennett1992communication}%
  \BibitemOpen
  \bibfield  {author} {\bibinfo {author} {\bibfnamefont {C.~H.}\ \bibnamefont
  {Bennett}}\ and\ \bibinfo {author} {\bibfnamefont {S.~J.}\ \bibnamefont
  {Wiesner}},\ }\href@noop {} {\bibfield  {journal} {\bibinfo  {journal} {Phys.
  Rev. Lett.}\ }\textbf {\bibinfo {volume} {69}},\ \bibinfo {pages} {2881}
  (\bibinfo {year} {1992})}\BibitemShut {NoStop}%
\bibitem [{\citenamefont {Braunstein}\ and\ \citenamefont
  {Kimble}(1998)}]{braunstein1998teleportation}%
  \BibitemOpen
  \bibfield  {author} {\bibinfo {author} {\bibfnamefont {S.~L.}\ \bibnamefont
  {Braunstein}}\ and\ \bibinfo {author} {\bibfnamefont {H.~J.}\ \bibnamefont
  {Kimble}},\ }\href@noop {} {\bibfield  {journal} {\bibinfo  {journal} {Phys.
  Rev. Lett.}\ }\textbf {\bibinfo {volume} {80}},\ \bibinfo {pages} {869}
  (\bibinfo {year} {1998})}\BibitemShut {NoStop}%
\bibitem [{\citenamefont {Abbott}\ \emph
  {et~al.}(2016{\natexlab{a}})\citenamefont {Abbott}, \citenamefont {Abbott},
  \citenamefont {Abbott}, \citenamefont {Abernathy}, \citenamefont {Acernese},
  \citenamefont {Ackley}, \citenamefont {Adams}, \citenamefont {Adams},
  \citenamefont {Addesso}, \citenamefont {Adhikari} \emph
  {et~al.}}]{abbott2016observation}%
  \BibitemOpen
  \bibfield  {author} {\bibinfo {author} {\bibfnamefont {B.~P.}\ \bibnamefont
  {Abbott}}, \bibinfo {author} {\bibfnamefont {R.}~\bibnamefont {Abbott}},
  \bibinfo {author} {\bibfnamefont {T.~D.}\ \bibnamefont {Abbott}}, \bibinfo
  {author} {\bibfnamefont {M.~R.}\ \bibnamefont {Abernathy}}, \bibinfo {author}
  {\bibfnamefont {F.}~\bibnamefont {Acernese}}, \bibinfo {author}
  {\bibfnamefont {K.}~\bibnamefont {Ackley}}, \bibinfo {author} {\bibfnamefont
  {C.}~\bibnamefont {Adams}}, \bibinfo {author} {\bibfnamefont
  {T.}~\bibnamefont {Adams}}, \bibinfo {author} {\bibfnamefont
  {P.}~\bibnamefont {Addesso}}, \bibinfo {author} {\bibfnamefont {R.~X.}\
  \bibnamefont {Adhikari}},  \emph {et~al.},\ }\href@noop {} {\bibfield
  {journal} {\bibinfo  {journal} {Phys. Rev. Lett.}\ }\textbf {\bibinfo
  {volume} {116}},\ \bibinfo {pages} {061102} (\bibinfo {year}
  {2016}{\natexlab{a}})}\BibitemShut {NoStop}%
\bibitem [{\citenamefont {Abbott}\ \emph
  {et~al.}(2016{\natexlab{b}})\citenamefont {Abbott}, \citenamefont {Abbott},
  \citenamefont {Abbott}, \citenamefont {Abernathy}, \citenamefont {Acernese},
  \citenamefont {Ackley}, \citenamefont {Adams}, \citenamefont {Adams},
  \citenamefont {Addesso}, \citenamefont {Adhikari} \emph
  {et~al.}}]{abbott2016gw151226}%
  \BibitemOpen
  \bibfield  {author} {\bibinfo {author} {\bibfnamefont {B.~P.}\ \bibnamefont
  {Abbott}}, \bibinfo {author} {\bibfnamefont {R.}~\bibnamefont {Abbott}},
  \bibinfo {author} {\bibfnamefont {T.~D.}\ \bibnamefont {Abbott}}, \bibinfo
  {author} {\bibfnamefont {M.~R.}\ \bibnamefont {Abernathy}}, \bibinfo {author}
  {\bibfnamefont {F.}~\bibnamefont {Acernese}}, \bibinfo {author}
  {\bibfnamefont {K.}~\bibnamefont {Ackley}}, \bibinfo {author} {\bibfnamefont
  {C.}~\bibnamefont {Adams}}, \bibinfo {author} {\bibfnamefont
  {T.}~\bibnamefont {Adams}}, \bibinfo {author} {\bibfnamefont
  {P.}~\bibnamefont {Addesso}}, \bibinfo {author} {\bibfnamefont {R.~X.}\
  \bibnamefont {Adhikari}},  \emph {et~al.},\ }\href@noop {} {\bibfield
  {journal} {\bibinfo  {journal} {Phys. Rev. Lett.}\ }\textbf {\bibinfo
  {volume} {116}},\ \bibinfo {pages} {241103} (\bibinfo {year}
  {2016}{\natexlab{b}})}\BibitemShut {NoStop}%
\bibitem [{\citenamefont {Miranowicz}\ and\ \citenamefont
  {Kielich}(1994)}]{miranowicz1994quantum}%
  \BibitemOpen
  \bibfield  {author} {\bibinfo {author} {\bibfnamefont {A.}~\bibnamefont
  {Miranowicz}}\ and\ \bibinfo {author} {\bibfnamefont {S.}~\bibnamefont
  {Kielich}},\ }\enquote {\bibinfo {title} {Quantum-statistical theory of
  {R}aman scattering processes},}\ in\ \href@noop {} {\emph {\bibinfo
  {booktitle} {Modern Nonlinear Optics}}},\ Vol.~\bibinfo {volume} {3}\
  (\bibinfo  {publisher} {John Wiley \& Sons, New York},\ \bibinfo {year}
  {1994})\ pp.\ \bibinfo {pages} {531--626}\BibitemShut {NoStop}%
\bibitem [{\citenamefont {Sen}\ and\ \citenamefont
  {Mandal}(2005)}]{sen2005squeezed}%
  \BibitemOpen
  \bibfield  {author} {\bibinfo {author} {\bibfnamefont {B.}~\bibnamefont
  {Sen}}\ and\ \bibinfo {author} {\bibfnamefont {S.}~\bibnamefont {Mandal}},\
  }\href@noop {} {\bibfield  {journal} {\bibinfo  {journal} {J. Mod. Opt.}\
  }\textbf {\bibinfo {volume} {52}},\ \bibinfo {pages} {1789} (\bibinfo {year}
  {2005})}\BibitemShut {NoStop}%
\bibitem [{\citenamefont {Sen}\ \emph {et~al.}(2007)\citenamefont {Sen},
  \citenamefont {Mandal},\ and\ \citenamefont {Pe{\v{r}}ina}}]{sen2007quantum}%
  \BibitemOpen
  \bibfield  {author} {\bibinfo {author} {\bibfnamefont {B.}~\bibnamefont
  {Sen}}, \bibinfo {author} {\bibfnamefont {S.}~\bibnamefont {Mandal}}, \ and\
  \bibinfo {author} {\bibfnamefont {J.}~\bibnamefont {Pe{\v{r}}ina}},\
  }\href@noop {} {\bibfield  {journal} {\bibinfo  {journal} {J. Phys. B}\
  }\textbf {\bibinfo {volume} {40}},\ \bibinfo {pages} {1417} (\bibinfo {year}
  {2007})}\BibitemShut {NoStop}%
\bibitem [{\citenamefont {Sen}\ and\ \citenamefont
  {Mandal}(2008)}]{sen2008amplitude}%
  \BibitemOpen
  \bibfield  {author} {\bibinfo {author} {\bibfnamefont {B.}~\bibnamefont
  {Sen}}\ and\ \bibinfo {author} {\bibfnamefont {S.}~\bibnamefont {Mandal}},\
  }\href@noop {} {\bibfield  {journal} {\bibinfo  {journal} {J. Mod. Opt.}\
  }\textbf {\bibinfo {volume} {55}},\ \bibinfo {pages} {1697} (\bibinfo {year}
  {2008})}\BibitemShut {NoStop}%
\bibitem [{\citenamefont {Sen}\ \emph {et~al.}(2011)\citenamefont {Sen},
  \citenamefont {Pe{\v{r}}inov{\'a}}, \citenamefont {Luk{\v{s}}}, \citenamefont
  {Pe{\v{r}}ina},\ and\ \citenamefont {K{\v{r}}epelka}}]{sen2011sub}%
  \BibitemOpen
  \bibfield  {author} {\bibinfo {author} {\bibfnamefont {B.}~\bibnamefont
  {Sen}}, \bibinfo {author} {\bibfnamefont {V.}~\bibnamefont
  {Pe{\v{r}}inov{\'a}}}, \bibinfo {author} {\bibfnamefont {A.}~\bibnamefont
  {Luk{\v{s}}}}, \bibinfo {author} {\bibfnamefont {J.}~\bibnamefont
  {Pe{\v{r}}ina}}, \ and\ \bibinfo {author} {\bibfnamefont {J.}~\bibnamefont
  {K{\v{r}}epelka}},\ }\href@noop {} {\bibfield  {journal} {\bibinfo  {journal}
  {J. Phys. B}\ }\textbf {\bibinfo {volume} {44}},\ \bibinfo {pages} {105503}
  (\bibinfo {year} {2011})}\BibitemShut {NoStop}%
\bibitem [{\citenamefont {Pathak}\ \emph {et~al.}(2013)\citenamefont {Pathak},
  \citenamefont {K{\u{r}}epelka},\ and\ \citenamefont
  {Pe{\v{r}}ina}}]{pathak2013nonclassicality}%
  \BibitemOpen
  \bibfield  {author} {\bibinfo {author} {\bibfnamefont {A.}~\bibnamefont
  {Pathak}}, \bibinfo {author} {\bibfnamefont {J.}~\bibnamefont
  {K{\u{r}}epelka}}, \ and\ \bibinfo {author} {\bibfnamefont {J.}~\bibnamefont
  {Pe{\v{r}}ina}},\ }\href@noop {} {\bibfield  {journal} {\bibinfo  {journal}
  {Phys. Lett. A}\ }\textbf {\bibinfo {volume} {377}},\ \bibinfo {pages} {2692}
  (\bibinfo {year} {2013})}\BibitemShut {NoStop}%
\bibitem [{\citenamefont {Sen}\ \emph {et~al.}(2013)\citenamefont {Sen},
  \citenamefont {Giri}, \citenamefont {Mandal}, \citenamefont {Ooi},\ and\
  \citenamefont {Pathak}}]{sen2013intermodal}%
  \BibitemOpen
  \bibfield  {author} {\bibinfo {author} {\bibfnamefont {B.}~\bibnamefont
  {Sen}}, \bibinfo {author} {\bibfnamefont {S.~K.}\ \bibnamefont {Giri}},
  \bibinfo {author} {\bibfnamefont {S.}~\bibnamefont {Mandal}}, \bibinfo
  {author} {\bibfnamefont {C.~H.~R.}\ \bibnamefont {Ooi}}, \ and\ \bibinfo
  {author} {\bibfnamefont {A.}~\bibnamefont {Pathak}},\ }\href@noop {}
  {\bibfield  {journal} {\bibinfo  {journal} {Phys. Rev. A}\ }\textbf {\bibinfo
  {volume} {87}},\ \bibinfo {pages} {022325} (\bibinfo {year}
  {2013})}\BibitemShut {NoStop}%
\bibitem [{\citenamefont {Giri}\ \emph {et~al.}(2016)\citenamefont {Giri},
  \citenamefont {Sen}, \citenamefont {Pathak},\ and\ \citenamefont
  {Jana}}]{giri2016higher}%
  \BibitemOpen
  \bibfield  {author} {\bibinfo {author} {\bibfnamefont {S.~K.}\ \bibnamefont
  {Giri}}, \bibinfo {author} {\bibfnamefont {B.}~\bibnamefont {Sen}}, \bibinfo
  {author} {\bibfnamefont {A.}~\bibnamefont {Pathak}}, \ and\ \bibinfo {author}
  {\bibfnamefont {P.~C.}\ \bibnamefont {Jana}},\ }\href@noop {} {\bibfield
  {journal} {\bibinfo  {journal} {Phys. Rev. A}\ }\textbf {\bibinfo {volume}
  {93}},\ \bibinfo {pages} {012340} (\bibinfo {year} {2016})}\BibitemShut
  {NoStop}%
\bibitem [{\citenamefont {Meekhof}\ \emph {et~al.}(1996)\citenamefont
  {Meekhof}, \citenamefont {Monroe}, \citenamefont {King}, \citenamefont
  {Itano},\ and\ \citenamefont {Wineland}}]{meekhof1996generation}%
  \BibitemOpen
  \bibfield  {author} {\bibinfo {author} {\bibfnamefont {D.}~\bibnamefont
  {Meekhof}}, \bibinfo {author} {\bibfnamefont {C.}~\bibnamefont {Monroe}},
  \bibinfo {author} {\bibfnamefont {B.}~\bibnamefont {King}}, \bibinfo {author}
  {\bibfnamefont {W.~M.}\ \bibnamefont {Itano}}, \ and\ \bibinfo {author}
  {\bibfnamefont {D.~J.}\ \bibnamefont {Wineland}},\ }\href@noop {} {\bibfield
  {journal} {\bibinfo  {journal} {Phys. Rev. Lett.}\ }\textbf {\bibinfo
  {volume} {76}},\ \bibinfo {pages} {1796} (\bibinfo {year}
  {1996})}\BibitemShut {NoStop}%
\bibitem [{\citenamefont {Chen}\ \emph {et~al.}(2006)\citenamefont {Chen},
  \citenamefont {Chen}, \citenamefont {Strassel}, \citenamefont {Yuan},
  \citenamefont {Zhao}, \citenamefont {Schmiedmayer},\ and\ \citenamefont
  {Pan}}]{chen2006deterministic}%
  \BibitemOpen
  \bibfield  {author} {\bibinfo {author} {\bibfnamefont {S.}~\bibnamefont
  {Chen}}, \bibinfo {author} {\bibfnamefont {Y.-A.}\ \bibnamefont {Chen}},
  \bibinfo {author} {\bibfnamefont {T.}~\bibnamefont {Strassel}}, \bibinfo
  {author} {\bibfnamefont {Z.-S.}\ \bibnamefont {Yuan}}, \bibinfo {author}
  {\bibfnamefont {B.}~\bibnamefont {Zhao}}, \bibinfo {author} {\bibfnamefont
  {J.}~\bibnamefont {Schmiedmayer}}, \ and\ \bibinfo {author} {\bibfnamefont
  {J.-W.}\ \bibnamefont {Pan}},\ }\href@noop {} {\bibfield  {journal} {\bibinfo
   {journal} {Phys. Rev. Lett.}\ }\textbf {\bibinfo {volume} {97}},\ \bibinfo
  {pages} {173004} (\bibinfo {year} {2006})}\BibitemShut {NoStop}%
\bibitem [{\citenamefont {Matsukevich}\ and\ \citenamefont
  {Kuzmich}(2004)}]{matsukevich2004quantum}%
  \BibitemOpen
  \bibfield  {author} {\bibinfo {author} {\bibfnamefont {D.}~\bibnamefont
  {Matsukevich}}\ and\ \bibinfo {author} {\bibfnamefont {A.}~\bibnamefont
  {Kuzmich}},\ }\href@noop {} {\bibfield  {journal} {\bibinfo  {journal}
  {Science}\ }\textbf {\bibinfo {volume} {306}},\ \bibinfo {pages} {663}
  (\bibinfo {year} {2004})}\BibitemShut {NoStop}%
\bibitem [{\citenamefont {Lee}\ \emph {et~al.}(2012)\citenamefont {Lee},
  \citenamefont {Sussman}, \citenamefont {Sprague}, \citenamefont
  {Michelberger}, \citenamefont {Reim}, \citenamefont {Nunn}, \citenamefont
  {Langford}, \citenamefont {Bustard}, \citenamefont {Jaksch},\ and\
  \citenamefont {Walmsley}}]{lee2012macroscopic}%
  \BibitemOpen
  \bibfield  {author} {\bibinfo {author} {\bibfnamefont {K.}~\bibnamefont
  {Lee}}, \bibinfo {author} {\bibfnamefont {B.}~\bibnamefont {Sussman}},
  \bibinfo {author} {\bibfnamefont {M.}~\bibnamefont {Sprague}}, \bibinfo
  {author} {\bibfnamefont {P.}~\bibnamefont {Michelberger}}, \bibinfo {author}
  {\bibfnamefont {K.}~\bibnamefont {Reim}}, \bibinfo {author} {\bibfnamefont
  {J.}~\bibnamefont {Nunn}}, \bibinfo {author} {\bibfnamefont {N.}~\bibnamefont
  {Langford}}, \bibinfo {author} {\bibfnamefont {P.}~\bibnamefont {Bustard}},
  \bibinfo {author} {\bibfnamefont {D.}~\bibnamefont {Jaksch}}, \ and\ \bibinfo
  {author} {\bibfnamefont {I.}~\bibnamefont {Walmsley}},\ }\href@noop {}
  {\bibfield  {journal} {\bibinfo  {journal} {Nature Photonics}\ }\textbf
  {\bibinfo {volume} {6}},\ \bibinfo {pages} {41} (\bibinfo {year}
  {2012})}\BibitemShut {NoStop}%
\bibitem [{\citenamefont {Kasperczyk}\ \emph {et~al.}(2016)\citenamefont
  {Kasperczyk}, \citenamefont {de~Aguiar~J{\'u}nior}, \citenamefont {Rabelo},
  \citenamefont {Saraiva}, \citenamefont {Santos}, \citenamefont {Novotny},\
  and\ \citenamefont {Jorio}}]{kasperczyk2016temporal}%
  \BibitemOpen
  \bibfield  {author} {\bibinfo {author} {\bibfnamefont {M.}~\bibnamefont
  {Kasperczyk}}, \bibinfo {author} {\bibfnamefont {F.~S.}\ \bibnamefont
  {de~Aguiar~J{\'u}nior}}, \bibinfo {author} {\bibfnamefont {C.}~\bibnamefont
  {Rabelo}}, \bibinfo {author} {\bibfnamefont {A.}~\bibnamefont {Saraiva}},
  \bibinfo {author} {\bibfnamefont {M.~F.}\ \bibnamefont {Santos}}, \bibinfo
  {author} {\bibfnamefont {L.}~\bibnamefont {Novotny}}, \ and\ \bibinfo
  {author} {\bibfnamefont {A.}~\bibnamefont {Jorio}},\ }\href@noop {}
  {\bibfield  {journal} {\bibinfo  {journal} {Phys. Rev. Lett.}\ }\textbf
  {\bibinfo {volume} {117}},\ \bibinfo {pages} {243603} (\bibinfo {year}
  {2016})}\BibitemShut {NoStop}%
\bibitem [{\citenamefont {Walls}(1970)}]{walls1970quantum}%
  \BibitemOpen
  \bibfield  {author} {\bibinfo {author} {\bibfnamefont {D.~F.}\ \bibnamefont
  {Walls}},\ }\href@noop {} {\bibfield  {journal} {\bibinfo  {journal}
  {Zeitschrift f{\"u}r Physik A Hadrons and Nuclei}\ }\textbf {\bibinfo
  {volume} {237}},\ \bibinfo {pages} {224} (\bibinfo {year}
  {1970})}\BibitemShut {NoStop}%
\bibitem [{\citenamefont {Duan}\ \emph {et~al.}(2001)\citenamefont {Duan},
  \citenamefont {Lukin}, \citenamefont {Cirac},\ and\ \citenamefont
  {Zoller}}]{duan2001long}%
  \BibitemOpen
  \bibfield  {author} {\bibinfo {author} {\bibfnamefont {L.-M.}\ \bibnamefont
  {Duan}}, \bibinfo {author} {\bibfnamefont {M.~D.}\ \bibnamefont {Lukin}},
  \bibinfo {author} {\bibfnamefont {J.~I.}\ \bibnamefont {Cirac}}, \ and\
  \bibinfo {author} {\bibfnamefont {P.}~\bibnamefont {Zoller}},\ }\href@noop {}
  {\bibfield  {journal} {\bibinfo  {journal} {Nature}\ }\textbf {\bibinfo
  {volume} {414}},\ \bibinfo {pages} {413} (\bibinfo {year}
  {2001})}\BibitemShut {NoStop}%
\bibitem [{\citenamefont {Kuzmich}\ \emph {et~al.}(2003)\citenamefont
  {Kuzmich}, \citenamefont {Bowen}, \citenamefont {Boozer}, \citenamefont
  {Boca}, \citenamefont {Chou}, \citenamefont {Duan},\ and\ \citenamefont
  {Kimble}}]{kuzmich2003generation}%
  \BibitemOpen
  \bibfield  {author} {\bibinfo {author} {\bibfnamefont {A.}~\bibnamefont
  {Kuzmich}}, \bibinfo {author} {\bibfnamefont {W.}~\bibnamefont {Bowen}},
  \bibinfo {author} {\bibfnamefont {A.}~\bibnamefont {Boozer}}, \bibinfo
  {author} {\bibfnamefont {A.}~\bibnamefont {Boca}}, \bibinfo {author}
  {\bibfnamefont {C.}~\bibnamefont {Chou}}, \bibinfo {author} {\bibfnamefont
  {L.-M.}\ \bibnamefont {Duan}}, \ and\ \bibinfo {author} {\bibfnamefont
  {H.}~\bibnamefont {Kimble}},\ }\href@noop {} {\bibfield  {journal} {\bibinfo
  {journal} {Nature}\ }\textbf {\bibinfo {volume} {423}},\ \bibinfo {pages}
  {731} (\bibinfo {year} {2003})}\BibitemShut {NoStop}%
\bibitem [{\citenamefont {Riedinger}\ \emph {et~al.}(2016)\citenamefont
  {Riedinger}, \citenamefont {Hong}, \citenamefont {Norte}, \citenamefont
  {Slater}, \citenamefont {Shang}, \citenamefont {Krause}, \citenamefont
  {Anant}, \citenamefont {Aspelmeyer},\ and\ \citenamefont
  {Gr{\"o}blacher}}]{riedinger2016non}%
  \BibitemOpen
  \bibfield  {author} {\bibinfo {author} {\bibfnamefont {R.}~\bibnamefont
  {Riedinger}}, \bibinfo {author} {\bibfnamefont {S.}~\bibnamefont {Hong}},
  \bibinfo {author} {\bibfnamefont {R.~A.}\ \bibnamefont {Norte}}, \bibinfo
  {author} {\bibfnamefont {J.~A.}\ \bibnamefont {Slater}}, \bibinfo {author}
  {\bibfnamefont {J.}~\bibnamefont {Shang}}, \bibinfo {author} {\bibfnamefont
  {A.~G.}\ \bibnamefont {Krause}}, \bibinfo {author} {\bibfnamefont
  {V.}~\bibnamefont {Anant}}, \bibinfo {author} {\bibfnamefont
  {M.}~\bibnamefont {Aspelmeyer}}, \ and\ \bibinfo {author} {\bibfnamefont
  {S.}~\bibnamefont {Gr{\"o}blacher}},\ }\href@noop {} {\bibfield  {journal}
  {\bibinfo  {journal} {Nature}\ }\textbf {\bibinfo {volume} {530}},\ \bibinfo
  {pages} {313} (\bibinfo {year} {2016})}\BibitemShut {NoStop}%
\bibitem [{\citenamefont {Dou}\ \emph {et~al.}(2018)\citenamefont {Dou},
  \citenamefont {Yang}, \citenamefont {Du}, \citenamefont {Lao}, \citenamefont
  {Gao}, \citenamefont {Qiao}, \citenamefont {Li}, \citenamefont {Pang},
  \citenamefont {Feng}, \citenamefont {Tang} \emph
  {et~al.}}]{dou2018broadband}%
  \BibitemOpen
  \bibfield  {author} {\bibinfo {author} {\bibfnamefont {J.-P.}\ \bibnamefont
  {Dou}}, \bibinfo {author} {\bibfnamefont {A.-l.}\ \bibnamefont {Yang}},
  \bibinfo {author} {\bibfnamefont {M.-Y.}\ \bibnamefont {Du}}, \bibinfo
  {author} {\bibfnamefont {D.}~\bibnamefont {Lao}}, \bibinfo {author}
  {\bibfnamefont {J.}~\bibnamefont {Gao}}, \bibinfo {author} {\bibfnamefont
  {L.-F.}\ \bibnamefont {Qiao}}, \bibinfo {author} {\bibfnamefont
  {H.}~\bibnamefont {Li}}, \bibinfo {author} {\bibfnamefont {X.-L.}\
  \bibnamefont {Pang}}, \bibinfo {author} {\bibfnamefont {Z.}~\bibnamefont
  {Feng}}, \bibinfo {author} {\bibfnamefont {H.}~\bibnamefont {Tang}},  \emph
  {et~al.},\ }\href@noop {} {\bibfield  {journal} {\bibinfo  {journal} {Comm.
  Phys.}\ }\textbf {\bibinfo {volume} {1}},\ \bibinfo {pages} {55} (\bibinfo
  {year} {2018})}\BibitemShut {NoStop}%
\bibitem [{\citenamefont {Ding}(2018)}]{ding2018raman}%
  \BibitemOpen
  \bibfield  {author} {\bibinfo {author} {\bibfnamefont {D.-S.}\ \bibnamefont
  {Ding}},\ }in\ \href@noop {} {\emph {\bibinfo {booktitle} {Broad Bandwidth
  and High Dimensional Quantum Memory Based on Atomic Ensembles}}}\ (\bibinfo
  {publisher} {Springer},\ \bibinfo {year} {2018})\ pp.\ \bibinfo {pages}
  {91--107}\BibitemShut {NoStop}%
\bibitem [{\citenamefont {Saraiva}\ \emph {et~al.}(2017)\citenamefont
  {Saraiva}, \citenamefont {de~Aguiar~J{\'u}nior}, \citenamefont {e~Souza},
  \citenamefont {Pena}, \citenamefont {Monken}, \citenamefont {Santos},
  \citenamefont {Koiller},\ and\ \citenamefont {Jorio}}]{saraiva2017photonic}%
  \BibitemOpen
  \bibfield  {author} {\bibinfo {author} {\bibfnamefont {A.}~\bibnamefont
  {Saraiva}}, \bibinfo {author} {\bibfnamefont {F.~S.}\ \bibnamefont
  {de~Aguiar~J{\'u}nior}}, \bibinfo {author} {\bibfnamefont {R.~d.~M.}\
  \bibnamefont {e~Souza}}, \bibinfo {author} {\bibfnamefont {A.~P.}\
  \bibnamefont {Pena}}, \bibinfo {author} {\bibfnamefont {C.~H.}\ \bibnamefont
  {Monken}}, \bibinfo {author} {\bibfnamefont {M.~F.}\ \bibnamefont {Santos}},
  \bibinfo {author} {\bibfnamefont {B.}~\bibnamefont {Koiller}}, \ and\
  \bibinfo {author} {\bibfnamefont {A.}~\bibnamefont {Jorio}},\ }\href@noop {}
  {\bibfield  {journal} {\bibinfo  {journal} {Phys. Rev. Lett.}\ }\textbf
  {\bibinfo {volume} {119}},\ \bibinfo {pages} {193603} (\bibinfo {year}
  {2017})}\BibitemShut {NoStop}%
\bibitem [{\citenamefont {Roelli}\ \emph {et~al.}(2016)\citenamefont {Roelli},
  \citenamefont {Galland}, \citenamefont {Piro},\ and\ \citenamefont
  {Kippenberg}}]{roelli2016molecular}%
  \BibitemOpen
  \bibfield  {author} {\bibinfo {author} {\bibfnamefont {P.}~\bibnamefont
  {Roelli}}, \bibinfo {author} {\bibfnamefont {C.}~\bibnamefont {Galland}},
  \bibinfo {author} {\bibfnamefont {N.}~\bibnamefont {Piro}}, \ and\ \bibinfo
  {author} {\bibfnamefont {T.~J.}\ \bibnamefont {Kippenberg}},\ }\href@noop {}
  {\bibfield  {journal} {\bibinfo  {journal} {Nature Nanotechnology}\ }\textbf
  {\bibinfo {volume} {11}},\ \bibinfo {pages} {164} (\bibinfo {year}
  {2016})}\BibitemShut {NoStop}%
\bibitem [{\citenamefont {Pieczonkov{\'a}}\ and\ \citenamefont
  {Pe{\v{r}}ina}(1981)}]{pieczonkova1981statistical}%
  \BibitemOpen
  \bibfield  {author} {\bibinfo {author} {\bibfnamefont {A.}~\bibnamefont
  {Pieczonkov{\'a}}}\ and\ \bibinfo {author} {\bibfnamefont {J.}~\bibnamefont
  {Pe{\v{r}}ina}},\ }\href@noop {} {\bibfield  {journal} {\bibinfo  {journal}
  {Czech. J. Phys. B}\ }\textbf {\bibinfo {volume} {31}},\ \bibinfo {pages}
  {837} (\bibinfo {year} {1981})}\BibitemShut {NoStop}%
\bibitem [{\citenamefont {Pe{\v{r}}inov{\'a}}\ \emph
  {et~al.}(1979)\citenamefont {Pe{\v{r}}inov{\'a}}, \citenamefont
  {Pe{\v{r}}ina}, \citenamefont {Szlachetka},\ and\ \citenamefont
  {Kielich}}]{perinova1979quantum}%
  \BibitemOpen
  \bibfield  {author} {\bibinfo {author} {\bibfnamefont {V.}~\bibnamefont
  {Pe{\v{r}}inov{\'a}}}, \bibinfo {author} {\bibfnamefont {J.}~\bibnamefont
  {Pe{\v{r}}ina}}, \bibinfo {author} {\bibfnamefont {P.}~\bibnamefont
  {Szlachetka}}, \ and\ \bibinfo {author} {\bibfnamefont {S.}~\bibnamefont
  {Kielich}},\ }\href@noop {} {\bibfield  {journal} {\bibinfo  {journal} {Acta
  Phys. Polonica A}\ }\textbf {\bibinfo {volume} {56}},\ \bibinfo {pages} {267}
  (\bibinfo {year} {1979})}\BibitemShut {NoStop}%
\bibitem [{\citenamefont {Aasi}\ \emph {et~al.}(2013)\citenamefont {Aasi},
  \citenamefont {Abadie}, \citenamefont {Abbott}, \citenamefont {Abbott},
  \citenamefont {Abbott}, \citenamefont {Abernathy}, \citenamefont {Adams},
  \citenamefont {Adams}, \citenamefont {Addesso}, \citenamefont {Adhikari}
  \emph {et~al.}}]{aasi2013enhanced}%
  \BibitemOpen
  \bibfield  {author} {\bibinfo {author} {\bibfnamefont {J.}~\bibnamefont
  {Aasi}}, \bibinfo {author} {\bibfnamefont {J.}~\bibnamefont {Abadie}},
  \bibinfo {author} {\bibfnamefont {B.~P.}\ \bibnamefont {Abbott}}, \bibinfo
  {author} {\bibfnamefont {R.}~\bibnamefont {Abbott}}, \bibinfo {author}
  {\bibfnamefont {T.~D.}\ \bibnamefont {Abbott}}, \bibinfo {author}
  {\bibfnamefont {M.~R.}\ \bibnamefont {Abernathy}}, \bibinfo {author}
  {\bibfnamefont {C.}~\bibnamefont {Adams}}, \bibinfo {author} {\bibfnamefont
  {T.}~\bibnamefont {Adams}}, \bibinfo {author} {\bibfnamefont
  {P.}~\bibnamefont {Addesso}}, \bibinfo {author} {\bibfnamefont {R.~X.}\
  \bibnamefont {Adhikari}},  \emph {et~al.},\ }\href@noop {} {\bibfield
  {journal} {\bibinfo  {journal} {Nature Photonics}\ }\textbf {\bibinfo
  {volume} {7}},\ \bibinfo {pages} {613} (\bibinfo {year} {2013})}\BibitemShut
  {NoStop}%
\bibitem [{\citenamefont {Grote}\ \emph {et~al.}(2013)\citenamefont {Grote},
  \citenamefont {Danzmann}, \citenamefont {Dooley}, \citenamefont {Schnabel},
  \citenamefont {Slutsky},\ and\ \citenamefont {Vahlbruch}}]{grote2013first}%
  \BibitemOpen
  \bibfield  {author} {\bibinfo {author} {\bibfnamefont {H.}~\bibnamefont
  {Grote}}, \bibinfo {author} {\bibfnamefont {K.}~\bibnamefont {Danzmann}},
  \bibinfo {author} {\bibfnamefont {K.~L.}\ \bibnamefont {Dooley}}, \bibinfo
  {author} {\bibfnamefont {R.}~\bibnamefont {Schnabel}}, \bibinfo {author}
  {\bibfnamefont {J.}~\bibnamefont {Slutsky}}, \ and\ \bibinfo {author}
  {\bibfnamefont {H.}~\bibnamefont {Vahlbruch}},\ }\href@noop {} {\bibfield
  {journal} {\bibinfo  {journal} {Phys. Rev. Lett.}\ }\textbf {\bibinfo
  {volume} {110}},\ \bibinfo {pages} {181101} (\bibinfo {year}
  {2013})}\BibitemShut {NoStop}%
\end{thebibliography}%

\appendix

\section*{Appendix A: Terms in the solution}

\setcounter{equation}{0} \renewcommand{\theequation}{A.\arabic{equation}}

The solutions reported in Eqs. (\ref{eq:Terms-in-CF}) and (\ref{eq:Terms-in-CF-Chaotic}),
which are obtained with the help of Eq. (\ref{eq:SM-soln}), contain
different functions defined as follows 
\begin{equation}
\begin{array}{lcl}
f_{1} & = & \exp\left(-i\omega_{L}t\right),\\
f_{2} & = & -\frac{g^{*}f_{1}}{\Delta\omega_{1}}\left[\exp\left(-i\Delta\omega_{1}t\right)-1\right],\\
f_{3} & = & \frac{\chi f_{1}}{\Delta\omega_{2}}\left[\exp\left(i\Delta\omega_{2}t\right)-1\right],\\
f_{4} & = & \frac{-\chi g^{*}f_{1}}{\Delta\omega_{2}}\left[\frac{\exp\left[-i\left(\Delta\omega_{1}-\Delta\omega_{2}\right)t\right]-1}{\Delta\omega_{1}-\Delta\omega_{2}}-\frac{\exp\left(-i\Delta\omega_{1}t\right)}{\Delta\omega_{1}}\right]\\
 & - & \frac{\chi g^{*}f_{1}}{\Delta\omega_{1}}\left[\frac{\exp\left[-i\left(\Delta\omega_{1}-\Delta\omega_{2}\right)t\right]-1}{\Delta\omega_{1}-\Delta\omega_{2}}+\frac{\exp\left(i\Delta\omega_{2}t\right)}{\Delta\omega_{2}}\right],\\
f_{5} & = & f_{6}\\
 & = & \frac{|g|^{2}f_{1}}{\Delta\omega_{1}^{2}}\left[\exp\left(-i\Delta\omega_{1}t\right)-1\right]+\frac{i|g|^{2}tf_{1}}{\Delta\omega_{1}},\\
f_{7} & = & -f_{8}\\
 & = & \frac{|\chi|^{2}f_{1}}{\Delta\omega_{2}^{2}}\left[\exp\left(i\Delta\omega_{2}t\right)-1\right]-\frac{i|\chi|^{2}tf_{1}}{\Delta\omega_{2}},
\end{array}\label{eq:solutions of f}
\end{equation}

\begin{equation}
\begin{array}{lcl}
g_{1} & = & \exp\left(-i\omega_{S}t\right),\\
g_{2} & = & \frac{gg_{1}}{\Delta\omega_{1}}\left[\exp\left(i\Delta\omega_{1}t\right)-1\right],\\
g_{3} & = & \frac{\chi^{*}gg_{1}}{\Delta\omega_{2}}\left[\frac{\exp\left[i\left(\Delta\omega_{1}-\Delta\omega_{2}\right)t\right]-1}{\Delta\omega_{1}-\Delta\omega_{2}}-\frac{\exp\left(i\Delta\omega_{1}t\right)-1}{\Delta\omega_{1}}\right],\\
g_{4} & = & \frac{\chi gg_{1}}{\Delta\omega_{2}}\left[\frac{\exp\left[i\left(\Delta\omega_{1}+\Delta\omega_{2}\right)t\right]-1}{\Delta\omega_{1}+\Delta\omega_{2}}-\frac{\exp\left(i\Delta\omega_{1}t\right)-1}{\Delta\omega_{1}}\right],\\
g_{5} & = & -g_{6}=\frac{|g|^{2}g_{1}}{\Delta\omega_{1}^{2}}\left[\exp\left(i\Delta\omega_{1}t\right)-1\right]-\frac{i|g|^{2}tg_{1}}{\Delta\omega_{1}},
\end{array}\label{eq:solutions of g}
\end{equation}
\begin{equation}
\begin{array}{lcl}
h_{1} & = & \exp\left(-i\omega_{V}t\right),\\
h_{2} & = & \frac{gh_{1}}{\Delta\omega_{1}}\left[\exp\left(i\Delta\omega_{1}t\right)-1\right],\\
h_{3} & = & \frac{\chi h_{1}}{\Delta\omega_{2}}\left[\exp\left(i\Delta\omega_{2}t\right)-1\right],\\
h_{4} & = & \frac{\chi gh_{1}}{\Delta\omega_{2}}\left[\frac{\exp\left[i\left(\Delta\omega_{1}+\Delta\omega_{2}\right)t\right]-1}{\Delta\omega_{1}+\Delta\omega_{2}}-\frac{\exp\left(i\Delta\omega_{1}t\right)}{\Delta\omega_{1}}\right]\\
 & - & \frac{\chi gh_{1}}{\Delta\omega_{1}}\left[\frac{\exp\left[i\left(\Delta\omega_{1}+\Delta\omega_{2}\right)t\right]-1}{\Delta\omega_{1}+\Delta\omega_{2}}-\frac{\exp\left(i\Delta\omega_{2}t\right)}{\Delta\omega_{2}}\right],\\
h_{5} & = & -h_{6}=-\frac{|g|^{2}h_{1}}{\Delta\omega_{1}^{2}}\left[\exp\left(i\Delta\omega_{1}t\right)-1\right]+\frac{i|g|^{2}th_{1}}{\Delta\omega_{1}},\\
h_{7} & = & -h_{8}=-\frac{|\chi|^{2}h_{1}}{\Delta\omega_{2}^{2}}\left[\exp\left(i\Delta\omega_{2}t\right)-1\right]+\frac{i|\chi|^{2}th_{1}}{\Delta\omega_{2}},
\end{array}\label{eq:eq:solutions of h}
\end{equation}
\begin{equation}
\begin{array}{lcl}
l_{1} & = & \exp\left(-i\omega_{A}t\right),\\
l_{2} & = & -\frac{\chi^{*}l_{1}}{\Delta\omega_{2}}\left[\exp\left(-i\Delta\omega_{2}t\right)-1\right],\\
l_{3} & = & \frac{\chi^{*}gl_{1}}{\Delta\omega_{1}}\left[\frac{\exp\left[i\left(\Delta\omega_{1}-\Delta\omega_{2}\right)t\right]-1}{\Delta\omega_{1}-\Delta\omega_{2}}+\frac{\exp\left(-i\Delta\omega_{2}t\right)-1}{\Delta\omega_{2}}\right],\\
l_{4} & = & \frac{\chi^{*}g^{*}l_{1}}{\Delta\omega_{1}}\left[\frac{\exp\left[-i\left(\Delta\omega_{1}+\Delta\omega_{2}\right)t\right]-1}{\Delta\omega_{1}+\Delta\omega_{2}}-\frac{\exp\left(-i\Delta\omega_{2}t\right)-1}{\Delta\omega_{2}}\right],\\
l_{5} & = & l_{6}=\frac{|\chi|^{2}l_{1}}{\Delta\omega_{2}^{2}}\left[\exp\left(-i\Delta\omega_{2}t\right)-1\right]+\frac{i|\chi|^{2}tl_{1}}{\Delta\omega_{2}},
\end{array}\label{eq:eq:solutions of l}
\end{equation}
where $\Delta\omega_{1}=(\omega_{S}+\omega_{V}-\omega_{L})$ and $\Delta\omega_{2}=(\omega_{L}+\omega_{V}-\omega_{A})$
are detuning in Stokes and anti-Stokes generation processes.
\end{document}